\documentclass[a4paper,11pt]{article}
\usepackage{jcappub} 
\usepackage{lineno}
\usepackage{hyperref}
\usepackage{subcaption}
\usepackage{xcolor}
\usepackage{soul}
\usepackage{booktabs}
\usepackage{siunitx}
\usepackage{multirow}
\definecolor{bittersweet}{rgb}{1.0, 0.44, 0.37}

\usepackage{comment}

\title{\boldmath Baryon Acoustic Oscillations in tomographic Angular Density and Redshift Fluctuations}

\newcommand*{\Euclid}{\textit{Euclid}\xspace}
\newcommand*{\Planck}{\\newcommand*{\Euclid}{\textit{Euclid}\xspace}
\newcommand*{\Planck}{\textit{Planck}\xspace}
\newcommand*{\Herschel}{\textit{Herschel}\xspace}
\newcommand*{\Fermi}{\textit{Fermi}\xspace}
\newcommand*{\Chandra}{\textit{Chandra}\xspace}
\newcommand*\XMMN{XMM-\textit{Newton}\xspace}
\newcommand*{\WMAP}{\textit{Wilkinson} Microwave Anisotropy Probe\xspace}
\newcommand*{\Spitzer}{\textit{Spitzer} Space Telescope\xspace}
\newcommand*{\HST}{\textit{Hubble} Space Telescope\xspace}textit{Planck}\xspace}
\newcommand*{\Herschel}{\textit{Herschel}\xspace}
\newcommand*{\Fermi}{\textit{Fermi}\xspace}
\newcommand*{\Chandra}{\textit{Chandra}\xspace}
\newcommand*\XMMN{XMM-\textit{Newton}\xspace}
\newcommand*{\WMAP}{\textit{Wilkinson} Microwave Anisotropy Probe\xspace}
\newcommand*{\Spitzer}{\textit{Spitzer} Space Telescope\xspace}
\newcommand*{\HST}{\textit{Hubble} Space Telescope\xspace}

\author[a,b]{Paula S. Ferreira,}
\author[b,c]{Carlos Hernández-Monteagudo,}
\author[a,d]{Ribamar R. R. Reis}
\affiliation[a]{Instituto de Física, Universidade Federal do Rio de Janeiro,\\
Av. Athos da Silveira Ramos, 149 - Cidade Universitária, Rio de Janeiro, Brasil}
\affiliation[b]{
Instituto de Astrofísica de Canarias (IAC),
\\
C/Vía Láctea, s/n, E-38205, La Laguna, Tenerife, Spain}
\affiliation[c]{Departamento de Astrofísica, Universidad de La Laguna,
\\
Avenida Francisco Sánchez, s/n, E-38205, La Laguna, Tenerife, Spain}
\affiliation[d]{Observatório do Valongo, Universidade Federal do Rio de Janeiro\\
Ladeira Pedro Antônio, 43, Centro,  Rio de Janeiro,
Brasil}
\emailAdd{psfer@pos.if.ufrj.br, chm@iac.es, ribamar@if.ufrj.br}

\abstract{
In this work we examine the baryon acoustic oscillations (BAO) in 2D angular and redshift space $\{\theta, \Delta z\}$, with $\Delta z$ denoting the redshift difference between two given angular shells. We thus work in the context of tomographic analyses of the large scale structure (LSS) where data are sliced in different redshift shells and constraints on Cosmology are extracted from the auto and cross-angular spectra of two different probes, namely the standard galaxy angular density fluctuations (ADF, or 2D clustering), and the galaxy angular redshift fluctuations (ARF). For these two observables we study by first time how the BAO peak arises in the $\{\theta, \Delta z\}$ plane. Despite being a weak feature (particularly for $\Delta z \neq 0$), a Fisher forecast analysis shows that, a priori, most of the information on cosmological and galaxy bias parameters is carried by the BAO features in shell auto- and cross-angular power spectra. The same study shows that a joint probe analysis (ADF+ARF) increases the Fisher determinant associated to cosmological parameters such as $H_0$ or the Dark Energy Chevallier-Polarski-Linder (CPL) parameters $\{w_0,w_a\}$ by at least an order of magnitude. We also study how the Fisher information on cosmological and galaxy bias-related parameters behaves under different redshift shell configurations: including cross-correlations to neighbour shells extending up to $(\Delta z)^{\rm tot}\sim 0.6$ ($(\Delta z)^{\rm tot}\sim 0.4$) for ADF (ARF) is required for Fisher information to converge. At the same time, configurations using narrow shell widths ($\sigma_z \leq  0.02$) preserve the cosmological information associated to peculiar velocities and typically yield Fisher determinants that are about two orders of magnitudes larger than for wider shell ($\sigma_z>0.02$) configurations. In the context of upcoming surveys of the LSS like {\it Euclid}, DESI, \textit{Roman}, J-PAS, LSST or CSST, these Fisher forecasts further motivate the tomographic use of pure angular anisotropies as an alternative approach to confront the cosmological predictions with observations, while providing a way to test consistency with standard 3D approaches to analyse LSS surveys. 
}

\begin{document}
\maketitle
\flushbottom
\raggedbottom 
\section{Introduction}
\label{sec:intro}

According to the standard $\Lambda$CDM model describing the physics of the early Universe, Baryon Acoustic Oscillations (BAO) are imprinted in the Large Scale Structure (LSS) as the last spherical sound waves propagating in the photon-baryon fluid prior to recombination ($z>1100$), \cite[e.g.,][]{zeldovich1969recombination,peebles1970primeval,hu2002cosmic}. The front of such waves formed an overdense spherical shell of about 150~Mpc radius (hereafter denoted as $r_s$) in which large-scale structure tends to form with higher probability. Such BAO scale was first seen in the angular anisotropy pattern of the Cosmic Microwave Background radiation (CMB) \cite{miller2002qmap,henrot2003archeops,archeops2005,jones2006measurement,hinshaw2013nine} and later detected in the spatial distribution of galaxies \cite{huetsi2007power,eisenstein2005detection}.

In this work we focus on an alternative approach to characterise the BAO scale in the LSS. Since it constitutes a spherical, 3D structure that arises when studying the clustering of matter probes like galaxies and quasars, such a scale can be measured along the line of sight and on the plane of the sky. In the former case this measurement is sensitive to the Hubble parameter $H(z)$, i.e., the expansion rate of the Universe at redshift $z$, while in the latter it measures a proxy for the transverse comoving distance to the redshift $z$ where the BAO scale is being measured. Provided that, according to General Relativity, distances in space time evolve as a function of its energetic content, BAO have become, together with CMB and supernova type Ia (SNIa) observations, a superb probe to characterise the different energy components in our Universe.

Over the last 20 years, BAO have been an essential tool to measure distance ratios throughout different cosmological epochs. The first detections by  \cite{huetsi2007power,eisenstein2005detection} in the Sloan Digital Sky Survey (SDSS) were improved in subsequent spectroscopic surveys like the SDSS extension BOSS \cite{BossRoss15,Boss17}, 6dF \cite{6dF}, or eBOSS \cite{eBoss1_20,eBoss2_20}.   
While precise and accurate redshift measurements are required to measure the radial projection of the BAO scale (so far it has only been measured in spectroscopic surveys), its transverse projection has been measured in spectroscopic \cite{carvalho2016baryon,carvalho2020transverse,angularBAO_2,menote2022baryon,Ferreira2024} or in photometric surveys with poorer redshift precision \cite[e.g.,][]{angularBAO_1,BAO_DES_2022,2024arXiv240210696D,hinton2016measuring}. 

In the future, surveys like the Legacy Survey of Space and Time (LSST) \cite{collaboration2012large}, the \textit{Euclid} Wide Survey \cite{scaramella2022euclid}, the \textit{Roman Space Telescope} \cite{eifler2021cosmology} and \textit{China’s Space Survey Telescope} (CSST) survey \cite{zhan2021wide,miao2024forecasting}, will increase the number of objects detected, either by spectroscopic or photometric redshifts. 
There exists an hybrid type of surveys, dubbed as {\it spectro-photometric} surveys, which carry a set of multiple ($N\sim 40-60$) narrow and medium-width optical filters that give rise to {\it pseudo-}spectrum for each object in the footprint: surveys like Classifying Objects by Medium-Band Observations (COMBO) \cite{wolf2003combo}, Advanced Large Homogeneous Area Medium Band Redshift Astronomical (ALHAMBRA) \citep{hurtado2016alhambra}, Survey for High-z Absorption Red and Dead Sources (SHARDS) \cite{perez2012shards}, Physics of the Accelerating Universe (PAU) \cite{tonello2019pau}, Javalambre Photometric Local Universe Survey (J-PLUS) \cite{cenarro2019j}, or the Javalambre-Physics of the Accelerated universe Astrophysical Survey (J-PAS) \cite{benitez2014j} are able to provide high accuracy photometric redshifts, which, as shown in \cite{chaves2018effect}, should keep some degree of sensitivity upon the radial projection of the BAO scale.

The standard approach to measuring the BAO scale in spectroscopic surveys is via the computation of the 3D 2-point statistics of the galaxy distribution, namely the 3D spatial correlation function $\xi(\mathbf{s})$, or the 3D power spectrum $P(\mathbf{k})$. Provided that angular and redshift coordinates of galaxies are projected into spatial coordinates under the redshift-to-distance conversion dictated by a fiducial cosmological model, and given that measured galaxy redshifts contain a peculiar velocity component (induced by the galaxy's radial peculiar motion), the computation of the two-point statistics is conducted in {\em redshift} space, and this motivates their projection on Legendre multipoles ($\xi_{\ell}(s)\propto \int d\mu\, \xi (\mathbf{s}) \,{\cal L}_{\ell} (\mu=\hat{n}\cdot\hat{s}) ; P_{\ell}(k)\propto \int d\mu\,P(\mathbf{k})\,{\cal L}_{\ell} (\mu=\hat{k}\cdot\hat{n})$, for $\ell=0,2,4$)\footnote{The unitary vector $\hat{s}_{12}$ lies along the vector connecting the two galaxies defining the galaxy pair in the correlation function, while $\hat{n}$ denotes the line of sight direction on the sky.}, whose amplitude for $\ell>0$ depends on the galaxies' radial peculiar velocity field, but whose shape keeps memory of the BAO scale.

BAO are regarded as a very robust cosmological tool which have, however, undergone severe scrutiny over the last 10 years due to apparently strong tensions arising in the measurements of key cosmological parameters like the Hubble constant $H_0$: indeed, its measurements using calibrators in the local universe \cite{riess2022local,freedman2019carnegie} seem to lie $\gtrsim 4-5~\sigma$ away from BAO-based estimations \cite[e.g.,][]{philcox2020combining}. Significant efforts have not yet aligned the two data sets (see \cite{verdeReview24} for a recent review on the topic). A possible candidate to solve the tension was presented by \cite{lee2024chicago} using the J-region asymptotic giant branch (JAGB)  from the James Webb Space Telescope (JWST), although the situation is far from settled. In addition to $H_0$, the cosmological parameter $S_8 \equiv \sigma_8 \sqrt{\Omega_m/0.3}$ also shows $2-3~\sigma$ discrepancies between LSS survey estimates and CMB angular anisotropy predictions \cite{asgari2021kids,2024arXiv240210696D}. 

On top of all this, further controversy has just arisen recently since the analysis of CMB data from the data release 6 of the Atacama Cosmology Telescope (ACT) \cite{calabrese2025atacama} is in apparent contradiction to the full shape of the power spectrum and BAO analysis of DESI year 1 data \cite{adame2024desi}. While the latter find $2-3~\sigma$ evidence for time evolution in the Dark Energy component, ACT fails to find any significant statistical evidence for it. The just released BAO analyses from DESI year 3 \cite{karim2025desi} are still favouring some degree of dynamics for Dark Energy, while further insight suggests that it is the inferred matter density parameter $\Omega_m$ which seem to be in tension between the CMB and the LSS data sets. This is interesting since there exists also some degree of tension between the $\Omega_m$ estimates derived from DES BAO and from DES supernovae (SNe), with a time-dependent Dark Energy component (at least partially) solving the tension \cite{DES_BAO_SNe,o2022revealing,colgain2024putting}.
A time-dependent DE equation of state $w(z)$ can reconcile $\Omega_m$ discrepancies across redshifts \cite{colgain2025much}, but tensions persist between DESI BGS (full-shape) and DES SNe at the same redshift \cite{colgain2024desi}. Notably, $w(z)$ and $H_0$ are anti-correlated: phantom DE ($w(z) < -1$) increases $H_0$, while $w(z) > -1$ decreases it, exacerbating the tension with SH0ES \cite{riess2022comprehensive}. 

In this context, finding alternative probes and methods to test the model and constrain cosmological parameters may result crucial in identifying and isolating systematics and confirming or discarding claims based upon standard analyses. As mentioned above, in this paper we investigate further the sensitivity of angular probes that work on direct observables like the galaxies' angular positions and their measured redshifts. On top of the standard 2D clustering (or angular density fluctuations, hereafter ADF), we consider the angular redshift fluctuations \cite{hernandez2021density} (hereafter ARF) as a new cosmological probe providing additional constraining power. The ARF consider the galaxies' redshift as an extra entry whose angular anisotropy field is shown to be determined by the underlying matter and radial peculiar velocity fields \cite{hernandez2021density}. This approach is close to that of \cite{Silveira2024}, who searched for a dipolar pattern in the measured redshifts of BOSS and eBOSS luminous red galaxies (LRGs) and quasars (QSOs), in a (successful) attempt to recover the observed dipole of the CMB intensity that is caused by the local (inhomogeneous) matter distribution.

While in \cite{legrand2021high} a first and simplistic Fisher analysis on the extra sensitivity to cosmological parameters provided by ARF when combined to ADF was presented, there was no particularised study of the BAO in angle space. In this work we build upon \cite{legrand2021high}, characterising the information carried by the BAO in the $\{\theta, \Delta z\}$ plane involving arbitrary correlations between shells placed at different redshifts ($\Delta z \neq 0$), and exploring the dependence of the information content on cosmological and bias-related parameters on varying redshift shell configurations (in terms of their central redshifts, widths, redshift spacing, and total redshift increment $(\Delta z)^{\rm tot}$ sampled when considering cross-correlation to a varying number of distinct redshift shells). 

This work is structured as follows. In Sect.~\ref{sec:obs} we introduce a generic description of (projected) cosmological observables defined in the 2D celestial sphere, that is particularised for ADF and ARF. In Sect.~\ref{sec:surveys} we describe the two models adopted for spectroscopic LSS surveys (one similar to the {\it Euclid} spectroscopic survey, another close to DESI), that sample complementary redshift ranges. In Sect.~\ref{sec:bao} we characterise the BAO in angle and redshift space $\{\theta, \Delta z\}$ as seen by the ADF and ARF, while in Sect.~\ref{sec:fisher} we introduce the Fisher matrix methodology used to describe the information content under different observational configurations. In Sect.~\ref{sec:results} we present our results, which are discussed in Sect.~\ref{sec:discussion}. This section also contains our conclusions.

Throughout this work we adopt a flat $\Lambda$CDM cosmology compliant with {\it Planck} 2018 observations \citep{aghanim2020planck}, for which $\Omega_m=0.321$, $\Omega_{\Lambda}=0.679$, $\Omega_b=0.047$ are the critical density parameters for total matter, $\Lambda$, and baryons, respectively, and with $h=0.67$, $n_S=0.965$, and $A_s=2\times 10^{-9}$ for the reduced Hubble parameter, and the index and amplitude of the scalar power spectrum, respectively.

\section{Observables}\label{sec:obs}

Any cosmological observable living on the 2D celestial sphere can be written as a line of sight (LOS) integral of some sources $S(z,\hat{\mathbf{n}})$ weighted under a given window function $W(z)$:
\begin{equation}
{\cal O} ( \hat{\mathbf{n}} ) = \int dz\,W(z)S(z,\hat{\mathbf{n}}),
\label{eq:obs1}
\end{equation}
where in this case we have preferred writing the LOS integral in terms of observed redshift $z$. We shall handle observables like the galaxy angular number density that are intensive, i.e., are defined in units of inverse solid angle. One can instead write this integral in terms of the comoving radial distance $r$,
\begin{equation}
{\cal O} ( \hat{\mathbf{n}} ) = \int dr\, W(r)\, S(r,\hat{\mathbf{n}}),
\label{eq:obs2}
\end{equation}
where, in general, the source function $S(r,\hat{\mathbf{n}})$ may depend on the radial peculiar velocity of tracers $\mathbf{v}(r, \hat{\mathbf{n}})\cdot \hat{\mathbf{n}}$, since our observable is a function of the tracers' observed redshift and, to leading order, the observed redshift contains the radial peculiar velocity of the sources, $z+1 = (1+z_H)(1 + \mathbf{v}\cdot \hat{\mathbf{n}}/c)$, with $z_H$ the cosmological redshift due to the Hubble drift. 

In linear cosmological perturbation theory, one can write the source function in terms of the linear perturbations of density, velocities, gravitational potentials, etc, which, in Fourier space, can all be made proportional to the linear matter density contrast at some fixed time $\delta^m_{\mathbf{k}}$:

\begin{equation}
{\cal O} ( \hat{\mathbf{n}} ) = \int \mathrm{d} r \,W(r)\,\int \frac{d\mathbf{k}}{(2\pi)^3} \exp{\biggl(-i\mathbf{k}\cdot (r\hat{\mathbf{n}})\biggr) }\,S(k,r,\hat{\mathbf{k}}\cdot\hat{\mathbf{n}})\, \delta^m_{\mathbf{k}}.
\label{eq:obs3}
\end{equation}

The leading physical processes in the source term coupling the density contrast field $\delta_{\mathbf{k}}^m$ with the observables can be written, in 3D Fourier space, as a function of $k$ and $\hat{\mathbf{k}}\cdot\hat{\mathbf{n}}$\footnote{Radial peculiar velocities contribute to the sources with a $\hat{\mathbf{k}}\cdot\hat{\mathbf{n}}$ dependence in Fourier space.}. In this equation, we have kept the $r$ dependence in the Fourier version of $S$ since in this case this variable carries the ``time dependence" of this function (which is equivalent to a look-back time and to the comoving distance $r$ to the source). According to most inflationary models of the early universe, primordial fluctuations are Gaussian distributed, and so are the emerging observable cosmological anisotropies at their earlier stages. Under this assumption, the observables are perfectly determined by their statistical second-order momentum, namely the angular correlation function or its harmonic transform, the angular power spectrum:

\begin{equation}
\langle {\cal O}_A ( \hat{\mathbf{n}}_1 ) {\cal O}_B ( \hat{\mathbf{n}}_2 ) \rangle = \sum_{\ell} \frac{2\ell+1}{4\pi} C^{A,B}_\ell \,P_{\ell}( \hat{\mathbf{n}}_1 \cdot \hat{\mathbf{n}}_2 ).
\label{eq:2-point1}
\end{equation}

In this equation, $P_{\ell}(x)$ denotes the $\ell$-th order Legendre polynomial. The isotropy and statistical homogeneity imposed by the Cosmological Principle are reflected
in the angular correlation function above via its exclusive dependence on the angle separating the two lines of sight, $\hat{\mathbf{n}}_1,\hat{\mathbf{n}}_2$. The angular power spectrum for the two observables, $C^{A,B}_\ell$, can be written in terms of the linear matter power spectrum $P_m(k)$ that relates to the 3D Fourier modes of the linear matter density contrast via $ \langle \delta^m_{\mathbf{k}} (\delta^m_{\mathbf{q}})^\star\rangle = (2\pi)^3 P_m(k) \delta^D(\mathbf{k}-\mathbf{q})$, with $\delta^D$ the Dirac delta: 

\begin{equation}
    C_\ell^{A,B}=\frac{2}{\pi}\int \mathrm{d} k \,k^2\,P_m(k)\Delta_\ell^A(k) \Delta_\ell^B(k),
\label{eq:APS1}
\end{equation}
and where $\Delta_\ell^{A,B}(k)$ are the transfer functions associated to each observable. In this work we are using the model of \cite{mead2021hmcode} that introduces non-linear corrections to the otherwise linear matter power spectrum. These corrections include the enhancement of power on small scales due to the collapse of dark matter into halos, and the smearing of the acoustic baryonic oscillations due to their non-linear evolution. If the source functions do not depend upon the line of sight (LOS) $\hat{\mathbf{n}}$, then a transfer function can be expressed as a simple projection of the density modes:
\begin{equation}
    \Delta_{\ell}(k) = \int_{0}^{r_{\rm far}} \mathrm{d}r\, W(r)\, S(k,r) j_{\ell} (kr),
    \label{eq:DeltaL}
\end{equation}
with $j_{\ell}(x)$ the spherical Bessel function of order $\ell$\footnote{If instead $S=S(k,r,\hat{\mathbf{k}}\cdot\hat{\mathbf{n}})$ such that the dependence on $\hat{\mathbf{k}}\cdot\hat{\mathbf{n}}$ is linear, then 
the transfer functions are expressed in terms of derivatives of the spherical Bessel functions $j^{'}_\ell(x)$, and such derivatives can be rewritten as a linear combination of spherical Bessel functions of order $\ell \pm 1$. This yields to expressions similar to Eq.~(\ref{eq:DeltaL}) with a set of re-defined source functions $S(k,r)$.}.\\

If $A=B$ in Eq.~(\ref{eq:2-point1}) then it yields the autocorrelation function (in terms of the auto power spectrum $C_{\ell}^{A,A}$ or $C_{\ell}^{B,B}$ of the $A$ (or $B$) observables), while if $A\neq B  $ then Eq.~(\ref{eq:2-point1}) provides the cross angular correlation function in terms of the cross angular power spectrum $C_{\ell}^{A,B}$.

We next briefly outline the source functions for the two observables under study in this work: the 2D-clustering or angular density fluctuations (ADF), and the angular redshift fluctuations (ARF), first introduced in \cite{hernandez2021density}. Eventually we shall also consider the combination of both observables, ADF+ARF.

\subsection{ADF}

The density contrast expresses the deviation of a given field with respect to its average. For a galaxy survey containing full information on the spatial position of each galaxy (for which its redshift is understood as its radial coordinate), the 3D density contrast is computed as 
\begin{equation}
    \delta_g^{3D}(z,\hat{\mathbf{n}}) = \frac{n_g(z,\hat{\mathbf{n}})-\bar{n}_g(z)}{\bar{n}_g(z)},
\end{equation}
where $\bar{n}_g(z)$ is the average number density of galaxies at redshift $z$. 

If instead we are looking at the number density of galaxies projected under a given redshift window $W^i(z,\sigma_z)$\footnote{For simplicity here we assume redshift windows are Gaussian functions determined by their centre and width $\sigma_z$}, the 2D density contrast reads as   
\begin{equation}\label{eq:def_adf}
    \delta^i_g(\hat{\mathbf{n}}) = \frac{1}{N^i_g}\int_{z=0}^\infty dz\,\frac{\mathrm{d}V_\Omega}{dz}  \, \bar{n}_g(z)\delta_g^{3D}(z,\hat{\mathbf{n}})W_i(z;\sigma_z),
\end{equation}
where $W_i$ is the Gaussian window function for the $i$-th redshift shell with width $\sigma_z$, $\mathbf{\hat{n}}$ is a unitary vector pointing to any given direction to the sky, and $\delta_g^{3D}(z,\hat{\mathbf{n}})$ is the 3D galaxy density contrast and $\frac{\mathrm{d}V_\Omega}{dz}= \frac{c}{H_0} \frac{(1+z)^2 \, D_A^2(z)}{E(z)}$. The latter quantity can be (approximately) written in terms of matter 3D density contrast $\delta_m$ and a redshift-dependent linear bias $b_g(z)$, $\delta_g^{3D}(z,\hat{\mathbf{n}}) = b_g(z) \delta_m (z,\hat{\mathbf{n}})$. Finally, d$V_{\Omega}$ expresses the volume element per unit of solid angle, and $N^i_g$ is the average number of galaxies in that same volume element,
\begin{equation}
N_g^i = \int_{z=0}^\infty dz\,\frac{\mathrm{d}V_\Omega}{dz} \, \bar{n}_g(z) W_i(z;\sigma_z).
\label{eq:ang_no_density}
\end{equation}
$N_g^i$ provides the number of galaxies per unit solid angle.

\subsection{ARF}

Angular Redshift Fluctuations (ARF) describe the deviations of the galaxies' redshifts with respect to an angular average $z_c$ after selecting galaxies under a given redshift shell $W(z)$. From any given galaxy survey, the ARF may be obtained under the $i$-th redshift window $W_i$ via

\begin{equation}
    \hat{\delta_z}^{i}(\mathbf{\hat{n}}) \equiv \frac{\sum_{j\in\mathbf{\hat{n}}}  (z_j-z_c) W_i(z_j;\sigma_z)}{\langle\sum_{j\in \mathbf{\hat{n}}} W_i(z_j;\sigma_z)\rangle_{\mathbf{\hat{n}}}}.
    \label{eq:arf_data}
\end{equation}

In this equation, $z_j$ denotes the observed redshift for the $j$-th galaxy belonging to a pixel along the sky direction $\mathbf{\hat{n}}$, and $z_c$ is the average redshift under the entire redshift shell $W_i$ given by, 
\begin{equation}
z_c=\frac{\sum_j z_j W_i(z_j;\sigma_z)}{\sum_j W_i(z_j;\sigma_z)}, 
\label{eq:z_c_data}
\end{equation}
where in this case the $j$-index runs over all galaxies in the footprint. The ensemble average in the denominator of Eq.~(\ref{eq:arf_data}) is effectively computed over the survey's footprint. Following the definition for the ADF given in the previous section, the model for the ARF under the $i$-th redshift shell can now be introduced as

\begin{align}\label{eq:def_arf} 
\delta_{z}^i (\hat{\mathbf{n}}) = \frac{1}{N^i_g}\int_{z=0}^{\infty} dz\,\frac{\mathrm{d}V_\Omega}{dz}  \left( z-z_c\right)\bar{n}_g(z)\left[ 1+b_g(z) \delta_m^{3D}(z,\mathbf{\hat{n}})\right] W_i(z;\sigma_z).
\end{align}

The average redshift $z_c$ under the shell can be then written as
\begin{equation}
z_c=\frac{1}{N^i_g}\int_{z=0}^{\infty} dz\,\frac{\mathrm{d}V_\Omega}{dz}\,  z \, \bar{n}_g(z) W_i(z;\sigma_z).
\label{eq:z_c}
\end{equation}
We revisit now the subtleties quoted above that relate to the definition of redshift used in Eqs.~\ref{eq:def_arf} and \ref{eq:def_adf}. The integral runs over the {\rm observed} redshift $z$, which is related to the Hubble drift redshift $z_H$ via 
\begin{equation}
    z = z_H+z_{pec}+z_{\Phi},
    \label{eq:z_obs}
\end{equation}
where $z_{\rm pec}$ denotes the peculiar redshift induced by the emitters' radial peculiar velocities $\mathbf{v}\cdot \hat{\mathbf{n}}$,
\begin{equation}
z_{\rm pec} = \mathbf{v}\cdot \hat{\mathbf{n}}\,/c\, (1+z_H) + {\cal O}[ (\mathbf{v}\cdot \hat{\mathbf{n}}\,/c)^2 ],
\label{eq:z_pec}
\end{equation}
with $c$ is the speed of light, and $z_{\Phi}$ relativistic and gravitational redshifts corrections described in \citep{lima2022relativistic}. Thus the observed redshift $z$ in a point of space-time depends not only on its radial distance to the emitter, but also on the emitter's local radial peculiar velocity and gravitational fields (although the latter will be ignored hereafter since they are much smaller in amplitude). In practice, this means that the source term of Eq.~\ref{eq:DeltaL} for {\em both} ADF and ARF contains a term that is proportional to the radial peculiar velocity, so both ADF and ARF will be sensitive to the matter density contrast and radial peculiar velocity field under the corresponding redshift window $W_i(z,\sigma_z)$. We refer to \cite{hernandez2021density} for further details, and simply highlight here the fact that source term corresponding to the radial velocity term usually dominates over the term sensitive to the matter density contrast for narrow shells, i.e., for low values of $\sigma_z$ ($\sigma_z\lesssim 0.02$), in the relatively low $\ell$ domain ($\ell \lesssim 50$).

In this work we shall be using the {\tt ARFCAMB} code\footnote{Code accessible upon request at \url{https://github.com/chmATiac/ARFCAMB} } \cite{lima2022relativistic}. This code is a modification of the {\tt CAMB} code \cite{lewis2011camb} that incorporates ARF as an additional cosmological observable and computes all possible auto- and cross-correlations with the other set of observables, namely intensity and E and B polarization modes of the CMB, its lensing convergence, and the ADF and weak lensing maps corresponding to set of redshift shells that constitutes the code user's input. 

\section{Survey configuration and fiducial model}\label{sec:surveys}

\begin{figure}[htbp]
    \centering
    \includegraphics[width=.9\textwidth]{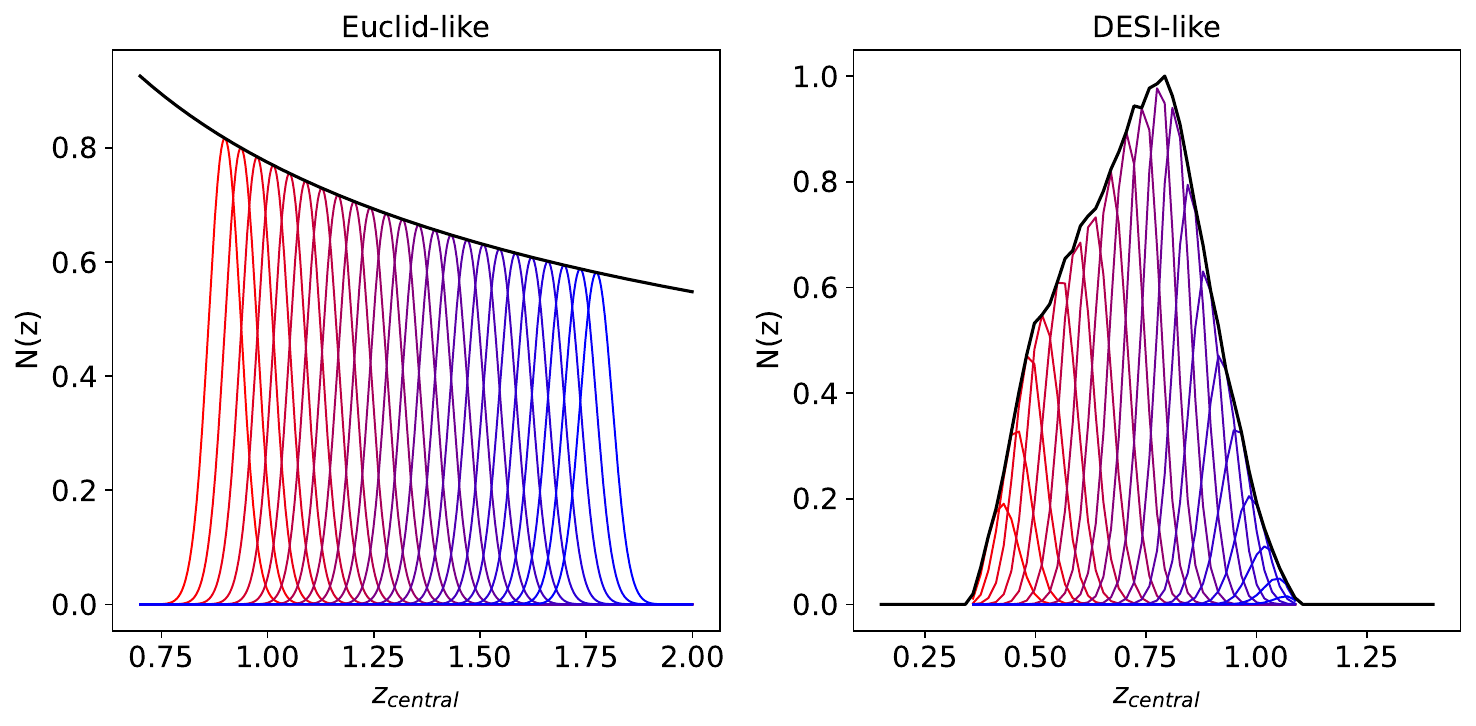}
    \caption{Redshift distributions adopted for the two models implemented (DESI-like model on the right panel, \textit{Euclid}-like model on the left one).}
    \label{fig:shells}
\end{figure}

For the Fisher matrix analysis we choose parametrizations close to the spectroscopic \textit{Euclid} \citep{blanchard2020euclid} and the DESI samples. The \textit{Euclid}-like survey is based on the H$\alpha$ slitless spectroscopic survey from \textit{Euclid}, covering the redshift range $z\in[0.9,1.8]$  with 1,950~gal~deg$^{-2}$, while the DESI-like survey redshift distribution is motivated by the Luminous Red Galaxies (LRG) sample from the Year 1 dataset \citep{zhou2023target}. The shape of the source number density versus redshift $N(z)$ follows DESI measurements on the Northern Galactic Cap, after applying a smoothing function on the original redshift distribution, and yielding an average galaxy angular density of $\sim$ 2,000~gal~deg$^{-2}$. In Table~\ref{tab:i}, we display the specifications for the two survey models. They have different redshift ranges and shapes, as can be seen in Fig.~\ref{fig:shells}. The choice for $\Delta z = \sigma_z = 0.038$ is related to the minimum redshift increment corresponding to the radial BAO in redshift units, which is approximately this choice for $z=0$, and increases steadily for higher redshifts\footnote{This is described in detail in section \ref{sec:bao_feat_sep}.}. 
\begin{table}[htbp]
\centering
\begin{tabular}{c|c|c}
\hline
&\it Euclid&DESI LRG\\
\hline
$f_{sky}$&36\%&34\%\\
$z$ range&$0.9<z<1.8$&$0.4<z<1.2$\\
$\sigma_z$ &0.038&0.038\\
$\Delta z$&0.038&0.038\\
$N(z)$ & $\sim 30\times10^6(z/0.6)^{-0.5}$ (as in \cite{legrand2021high}) & Smoothed from \cite{zhou2023target}\\
$\sigma_{ \rm Err}$ & 0.001 (1+z) \cite{nisp2025} &0.0005 (1+z) \cite{desi2024_val}\\
\hline
\end{tabular}
\caption{Configuration of the two survey models.}
\label{tab:i}
\end{table}

Our data vector will consist in a set of angular power spectra $\{C_\ell\}$ corresponding to a set of redshift shells and observables (ADF, ARF, or ADF+ARF). We may consider (auto-) angular power spectra involving only the same redshift shell or we may consider instead the (cross-) angular power spectra involving different redshift shells. Likewise, we may also build the $C_\ell$ vector upon the same observables (ADF$\times$ADF, ARF$\times$ARF), or we may consider both observables (ADF$\times$ADF {\em and}  ARF$\times$ARF) for our redshift shells. For $i$ and $j$ denoting redshift shells, the covariance matrix of the corresponding angular power spectrum is modelled via a Gaussian approximation as
\begin{equation}
\mathrm{CovM}_{i,j} \equiv \langle C_\ell^{i,j}C_{\ell'}^{i,j} \rangle - \langle C_\ell^{i,j} \rangle \langle C_{\ell'}^{i,j}\rangle= \delta^K_{\ell,\ell'}
    \frac{C_\ell^{i,i}C_{\ell}^{j,j} + (C_\ell^{i,j})^2 }{(2\ell+1)f_{\rm sky}},
\label{eq:covM0}
\end{equation}

where $\delta^K_{\ell,\ell'}$ is the Kronecker delta (equal to unity if $\ell=\ell'$ and zero otherwise), and $f_{\rm sky}$ is the fraction of the sky covered by the survey.

\begin{figure}
    \centering
    \begin{subfigure}[b]{0.4\textwidth}
        \centering
        \caption{ADF}
        \includegraphics[width=\textwidth]{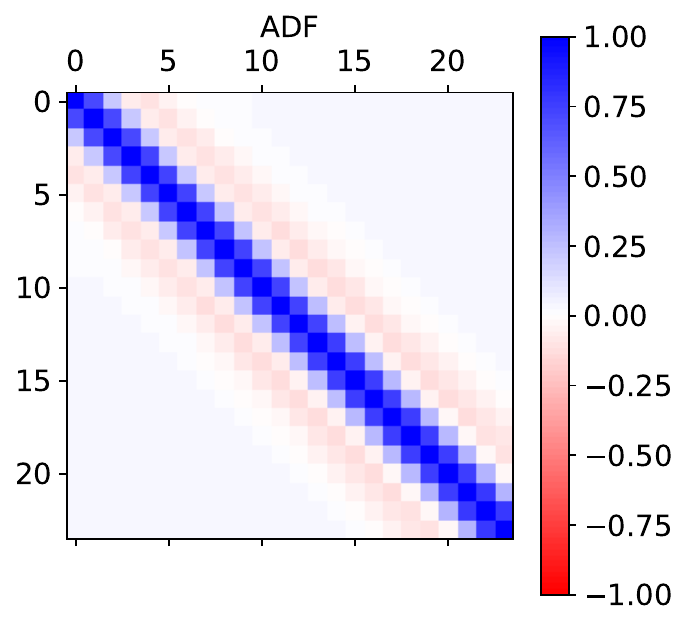}
        \label{fig:cov_matrix}
    \end{subfigure}
    \begin{subfigure}[b]{0.4\textwidth}
        \centering
        \caption{ARF}
        \includegraphics[width=\textwidth]{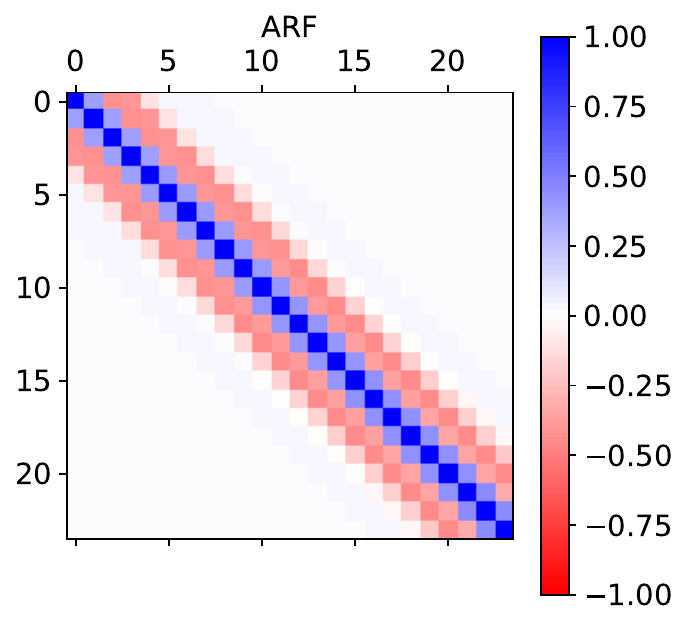}      
        \label{fig:cov_matrix2}
    \end{subfigure}
     \hfill
    \begin{subfigure}[b]{0.4\textwidth}
        \centering
        \caption{ADF+ARF}
        \includegraphics[width=\textwidth]{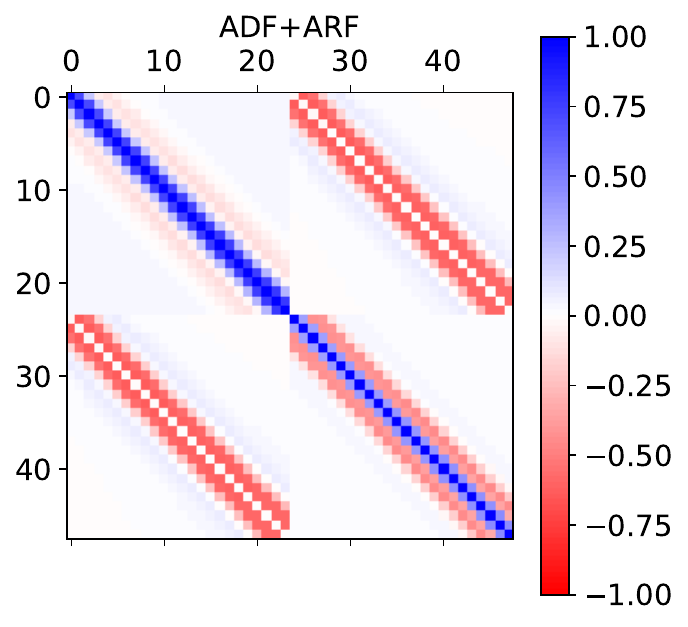}
        \label{cov_matrix3}
    \end{subfigure}
    \caption{Correlation matrix between shells from (a) ADF, (b) ARF, and (c) ADF+ARF at $\ell=20$ from Eq.~(\ref{eq:cov_mat}). Blue indicates positive correlation, white indicates no correlation, and red negative correlation.} 
    \label{fig:combined_cov_matrix}
\end{figure}

In Fig.~\ref{fig:combined_cov_matrix} we display correlation matrices\footnote{Correlation matrix is defined here as the covariance matrix of Eq.~\ref{eq:covM0} divided by the product of the square root of the corresponding diagonal elements, i.e., $\mathrm{CorrM}_{i,j}=\mathrm{CovM}_{i,j}/\sqrt{\mathrm{CovM}_{i,i}\mathrm{CovM}_{j,j}}$.} for the different shells under configuration, under the simplest case when no cross-correlation to neighbouring shells are included in the same data vector. Panels (a), (b), and (c) display the ADF, ARF, and ADF+ARF cases. Panel (c) shows practically null ADF$\times$ARF cross correlation in the diagonal of the off-diagonal panels, with some degree of anti-correlation in the nearest neighbour shells. 

We adopt a fiducial model based on {\it Planck} TT, TE, EE + lowE 2018 results \citep{aghanim2020planck}. The fiducial values for the cosmological parameters of relevance, together with the differentiation step adopted in the Fisher analysis below are given in Table~\ref{tab:cosmo_params}. We employ the Chevallier-Polarski-Linder (CPL) framework for parametrizing the Dark Energy equation of state, which is characterised by the parameters $w_0$ and $w_a$ that quantify the time evolution of the Dark Energy equation of state as $w(a)=w_0 + (1-a)\, w_a$, and this allows for an examination of the evolution of Dark Energy properties over cosmological timescales \cite{chevallier2001accelerating,linder2003exploring}.

Finally, regarding the galaxy bias, we chose a redshift-dependent bias $b(z)$ following a two-parameter model, $\tau$ and $\phi$: 
\begin{equation}
    b(z)=\tau(1+z)^{\phi}.
\end{equation}
The fiducial values for the two parameters are set to $\tau=1$ and $\phi=0.5$.  This is a relatively simplistic description of the linear bias that should capture its redshift evolution in a restricted redshift range. There are obvious possible improvements over this model, like a third parameter accounting for curvature in the redshift dependence, or some parameter(s) accounting for a scale-dependent, non-linear contribution. This (more realistic) bias modelling goes beyond the scope of this study, and is deferred to future, more realistic work involving data analysis.


\begin{table}[htpb]
    \centering
    \begin{tabular}{c|c|c}
        Parameter & Fiducial choice & Differentiation step\\
        \hline
         $H_0$(km s$^{-1}$ Mpc$^{-1}$)&67.27 & 0.67  \\
         $\Omega_b h^{2}$ & 0.02236& 0.0002\\
         $\Omega_c h^{2}$& 0.1202 & 0.001\\
         $w_0$& -1& 0.01\\
         $w_a$&0&0.01\\
         $n_s$&0.9649& -\\
         $A_s$&2.092$\times 10^{-9}$&-\\
         \hline$
         \tau$& 1&0.01 \\
         $\phi$& 0.5&0.005 \\
    \end{tabular}
    \caption{{\it Left column:} Fiducial cosmological parameters adopted from {\it Planck} 2018 \citep{aghanim2020planck} and our bias model. {\it Right column:} Corresponding differentiation steps for those parameters included in the Fisher analysis.}
    \label{tab:cosmo_params}
\end{table}

\section{BAOs seen in angular and redshift space}
\label{sec:bao}

As already described in the Introduction, the so-called ``BAO scale" corresponds to the sound horizon at the epoch of recombination ($z_{\rm rec}\sim 1,100$ \cite{zeldovich1969recombination}). Up to that epoch, Thomson's optical depth was high enough to keep radiation and baryons dynamically coupled. This coupling acted as a forced oscillator generating sound waves. As electrons recombined protons to form neutral hydrogen, this coupling vanished and CMB photons could propagate freely. The largest wave-front of the oscillations was frozen in the LSS with a radius that corresponds to the comoving sound horizon at that epoch, given by 
\begin{equation}
r_s = \int_0^{t=t_{\rm rec}} dt'\, \frac{c_s(t')}{a(t')},
\label{eq:sound_horizon}
\end{equation}
where $t_{\rm rec}$ is the time at $z=z_{\rm rec}$ and $c_s (t)$ is the time-dependent sound speed given by \cite{peebles1970primeval,hu1995effect}
\begin{equation}
c_s = \frac{c}{\sqrt{3(1+\frac{4\rho_b}{3\rho_{\gamma}})}}.
\label{eq:c_s}
\end{equation}
In this equations, $c$ denotes the speed of light, and $\rho_b$ and $\rho_{\gamma}$ correspond to the (time-dependent) baryon and radiation cosmic energy densities (the latter being exquisitely measured via CMB observations, \cite{hu1995cobe, aghanim2020planck}). 
The $r_s$ scale is also known as the BAO scale and can be expressed on the grounds of relatively simple physics driving the process of hydrogen recombination. From the observational point of view, the angular on sky projection of $r_s$ at $z=z_{\rm rec}$ (dubbed as $\theta_\star$) has been measured extremely accurately by the Planck Collaboration \cite{aghanim2020planck}: $100 \, \theta_{\star}=1.0411\pm 0.0003$, and after marginalization over other cosmological parameters, such constraint results into a measurement of $r_s=147.09 \pm 0.26$~Mpc.

The BAO scale thus marks the position of a spherical overdensity centred upon any other, {\it initial} overdensity. It shows up as a secondary peak at $r=r_s$ in the spatial correlation function of the galaxy field, defined as
\begin{equation}
\xi(r) \equiv \langle \delta_g(\mathbf{x}+\mathbf{r}) 
    \delta_g(\mathbf{x} ) \rangle_{\mathbf{x}} = 
    \frac{1}{(2\pi)^3}\int d\mathbf{k}\,
    P_{g,g}(k) \exp{[-i\mathbf{k}\cdot\mathbf{r}]},
\label{eq:xi_def}
\end{equation}
where the ensemble average takes place over all space $\mathbf{x}$ and $P_{g,g}(k)$ denotes the galaxy 3D power spectrum. 
This supposedly well-known scale (or ``standard ruler") $r_s$ has been extensively used in the last 20 years by LSS surveys to pin down properties of dark energy \citep[][to quote just a few]{eisenstein2005detection,adame2024desi,karim2025desi,raichoor2021completed}. These surveys convert (via a fiducial model) the angular and redshift information of galaxies into 3D comoving spatial coordinates. By imposing sphericity (or isotropy) of the BAO scale it is possible to constrain $\Omega_m$ without any assumption on the actual value of $r_s$ \cite{alcock1979evolution}, correct its fiducial value \cite{carter2019impact}, while it is also possible to set further constraints on $\Omega_b/\Omega_c$ and $H_0$ \cite{krolewski2024new,krolewski_prl}.

In this work we opt to investigate a different approach. We stick to observed quantities (galaxies' angular and redshift coordinates) to produce 2D, celestial maps of ADF and ARF computed under redshift shells centred upon different, successive redshifts, in order to conduct BAO tomography in angle and redshift space. 
We thus present here a work more focused on BAO than in \citep[Legrand et al.,][]{legrand2021high}, where a generic study on the extra information provided by the ARF on top of the ADF was presented for fixed {\it Euclid}- and DESI-like configurations. Here we explore in detail the BAO feature in the $\{\theta,z\}$ plane for both ADF and ARF, together with the dependence of the BAO information content on the shell widths, central redshifts, redshift spacing, and extent of cross-correlations to other shells.

\subsection{The BAO scale along the redshift axis (radial direction)}\label{sec:bao_feat_sep}

The BAO scale corresponds to the extent of a spherical sound wave that can be inferred statistically from LSS surveys. In a flat Friedmann–Lemaître–Robertson–Walker (FLRW) metric, for an observer placed at redshift $\tilde{z}$, a comoving radial distance  $r_s$ corresponds to a redshift increment $(\Delta z)^{BAO}$ in

\begin{equation}\label{eq:comoving}
    r_s= cH_0^{-1}\int_{\tilde{z}}^{\tilde{z}+(\Delta z)^{BAO}}\frac{dz'}{E(z')},
\end{equation}

with $E(z)\equiv\sqrt{\Omega_m (1+z)^3 + \Omega_{\rm DE}(z)}$, where $\Omega_m=\rho_m(z)/\rho_{\rm crit}(z)$ is the matter critical density parameter, and $\Omega_{\rm DE}$ is the corresponding dark energy density parameter given by $\rho_{\text{DE}}(z)/\rho_{\text{crit}}(z)$. The Dark Energy density $\rho_{\text{DE}}(z)$ can be expressed using the formula $\rho_{\text{DE},0} \, (1+z)^{3(1+w(z))}$, where $\rho_{\text{DE},0}$ denotes its present-day value and $w(z)$ its redshift or time dependent equation of state parameter.  The critical density of the universe $\rho_{\text{crit}}$ is expressed in terms of Newton's gravitational constant $G$ and the Hubble parameter $H(z)$ as $\rho_{\text{crit}}(z)=3H^2(z)/8\pi G$. The Dark Energy equation of state parameter $w(z)$ is modelled by CPL as $w(z)=w_0 + w_a[1-1/(1+z)]$. This is just one popular parametrization describing the cosmological evolution of $w(z)$ with its present day value, $w_0$, and one extra parameter, $w_a$, accounting for its time dependence.

At $\tilde{z}=0$ we find that $(\Delta z)^{BAO} \simeq 0.034$ for $r_s\simeq 100$~Mpc~$h^{-1}$, with an increasing trend for higher values of $\tilde{z}$. In order to be sensitive to the purely radial projection of $r_s$, the redshift separation between two different redshift shells must thus be below $(\Delta z)^{BAO} (\tilde{z})$. 

The width of the Gaussian shells under consideration ($\sigma_z$) plays also some role when comparing it to $(\Delta z)^{BAO} (\tilde{z})$: obviously $\sigma_z\ll (\Delta z)^{BAO} (\tilde{z})$ must hold if one attempts to unveil $r_s$ by cross-correlating different redshift shells. In this scenario, as it will be shown below, the BAO feature for ARF will not show up in the auto power spectrum/correlation function, contrary to ADF, for which BAO will show up even if $\sigma_z\ll (\Delta z)^{BAO} (\tilde{z})$. Instead, if $\sigma_z$ is comparable to $(\Delta z)^{BAO} (\tilde{z})$, the BAO scale will show up in the auto power spectrum or auto correlation function for both ADF and ARF. As $\sigma_z$ grows larger than $(\Delta z)^{BAO} (\tilde{z})$ the sensitivity to $r_s$ slowly declines. However, BAO do leave some signature at zero-lag in angle ($\theta=0$) when computing the cross-correlation function between two different, narrow enough redshift shells that lie at a redshift difference close to $(\Delta z)^{BAO} (\tilde{z})$. We  describe all these dependencies more formally in what follows below. 

It is worth noting that the redshift scale imposed by $(\Delta z)^{BAO} (\tilde{z})$ connects to the error limit adopted by the so-called spectro-photometric surveys targetting the radial BAO feature \cite{benitez2009optimal,benitez2014j}, typically set at the level of $\Delta z / (1+z) = 0.003$. In our work, the preliminary computation of $(\Delta z)^{BAO} (\tilde{z})$ above motivates the choice for the redshift shell widths and separation shown in Fig.~\ref{fig:shells} and given in Table~\ref{tab:i}. 

\subsection{The BAO scale along the angle axis (transversal direction)}

When looking at one particular redshift shell, the BAO scale will be projected along the transverse direction, i.e., along the direction perpendicular to the line of sight. If one computes the ADF angular correlation function in such shell, 
\begin{equation}
w_{g,g} (\theta )= \langle \delta_g(\hat{\mathbf{n}}) \delta_g(\hat{\mathbf{m}})\rangle_{\hat{\mathbf{m}}\cdot\hat{\mathbf{n}}=\cos\theta},
\label{eq:w_theta_def1}
\end{equation}
it should be equivalent to a projection of the 3D galaxy-galaxy correlation function $\xi(r)$ in Eq.~(\ref{eq:xi_def}) under that given redshift shell. Following the definition of the angular correlation function introduced in Sect.~\ref{sec:obs}, Eq.~(\ref{eq:2-point1}), we can write the angular correlation function for either ADF or ARF in terms of the corresponding angular power spectrum $C_\ell$:
\begin{equation}
    w(\theta) = \sum_\ell \frac{(2\ell+1)}{4 \pi } C_\ell P_\ell(\cos{\theta}),
\end{equation}
where $P_\ell$ is the Legendre polynomial of order $\ell$.

\begin{figure}[htpb]
    \centering
    \includegraphics[width=\textwidth]{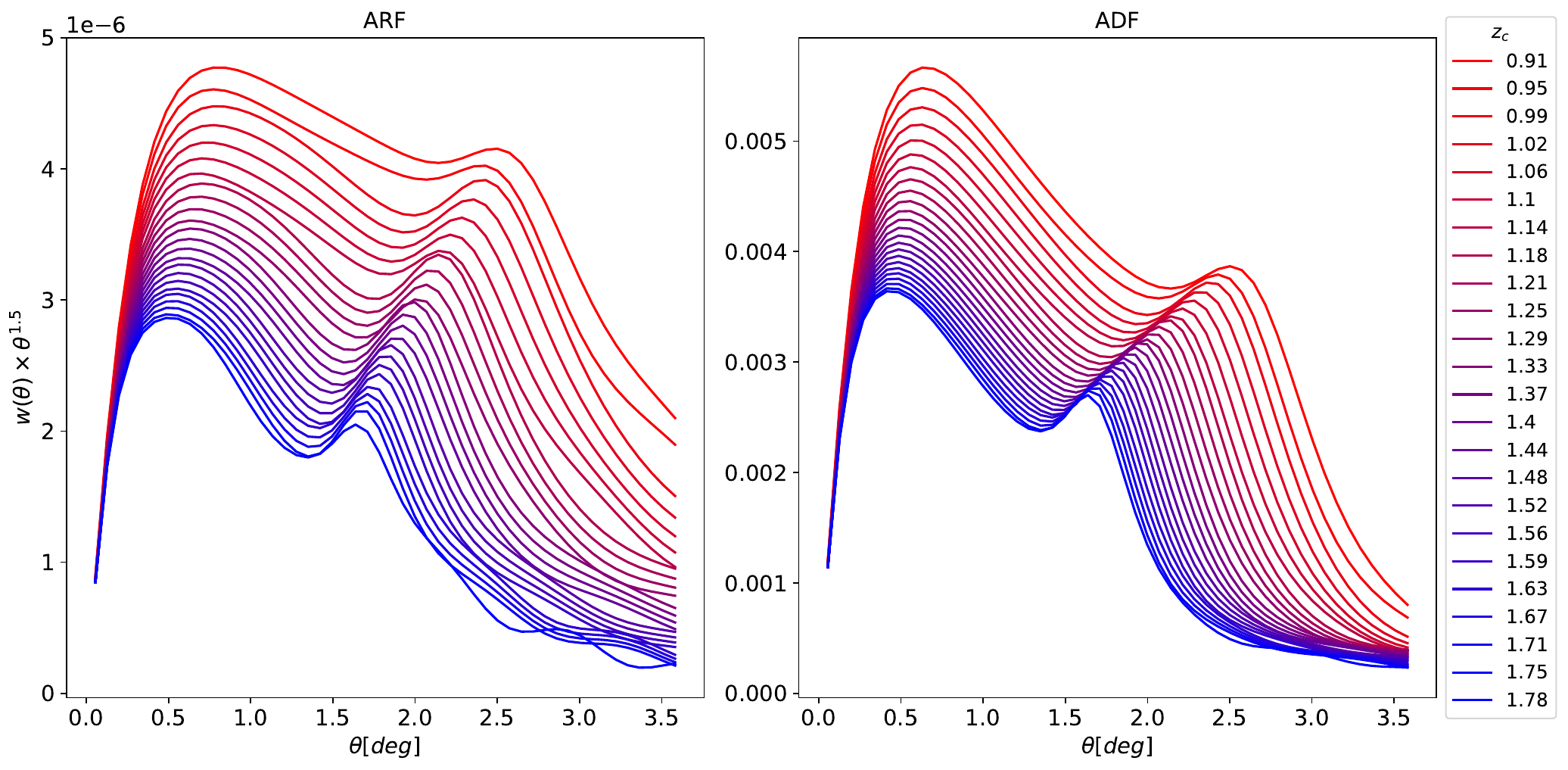}
    \caption{$w(\theta)$ for ARF (\textit{left panel}) and ADF (\textit{right panel}) for an \textit{Euclid}-like survey. The colours from red to blue increase with $z_c$, $\Delta z = \sigma_z=0.038$.}
    \label{fig:w_theta_examples}
\end{figure}

In Fig.~\ref{fig:w_theta_examples}, we show the angular auto-correlation function obtained from the two observables, ARF  (left panel) and ADF (right panel) for shell widths of $\sigma_z= 0.038$, after re-scaling by a $\theta^{1.5}$ factor that enhances the BAO feature at $\theta_{BAO}\sim 1.5-2.5$~deg. The colours represent the central redshift of the shell from lower $z$ (red) to higher $z$ (blue). ADF is a factor of $\sim 10^3$ larger than ARF, as expected from the $(z-z_c)$ kernel present in the ARF definition.  
Both observables display the BAO feature close to the expected position according to the equation:
\begin{equation}\label{eq:theta_bao}
    \theta_{BAO}=\frac{r_s}{(1+z_c) D_A(z_c)},
\end{equation}
where $D_A$ is the angular diameter distance. It can also be clearly seen that $\theta_{BAO}$ decreases with redshift. 

We next study how the relative height of the BAO peak w.r.t. the zero-lag peak ($\theta=0$) of the angular correlation depends upon the width $\sigma_z$ of the redshift shell. Intuitively one would expect that wider shells involve larger line-of-sight damping of scales smaller than the shell width. Those small scales typically contribute more to the primary (zero lag, $\theta=0$) peak of the correlation function than to the BAO peak (since the latter is built on scales around $r_s$). Thus it is expected that under narrower shells small-scale anisotropy contributes more to the zero-lag peak than to the BAO peak, and the ratio of the latter over the former should decrease for thinner shells. For ARF, instead, one has to account for the extra factor $(z-z_c)$ present in the kernel of Eq.~(\ref{eq:def_arf}). This factor effectively applies a radial/redshift gradient, such that if the density field is largely constant under the redshift shell, the amplitude of the ARF will be heavily suppressed. For the particular case of the BAO feature, if the shell width is much smaller than $(\Delta z)^{BAO}$ in Eq.~(\ref{eq:comoving}), then the BAO feature will not vary under the shell, and its projection under the ARF kernel will be small. 

\begin{figure}[htbp]
    \centering
    \begin{subfigure}[b]{0.45\textwidth}
        \centering
        \includegraphics[width=\textwidth]{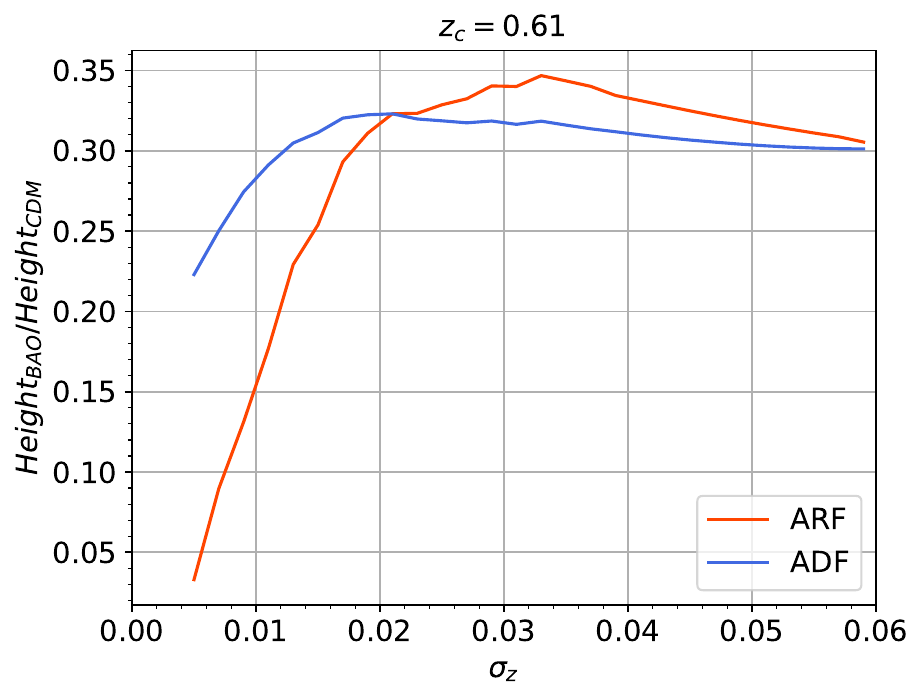}
        \label{fig:rel_heights2}
    \end{subfigure}
    \hfill
    \begin{subfigure}[b]{0.45\textwidth}
        \centering
        \includegraphics[width=\textwidth]{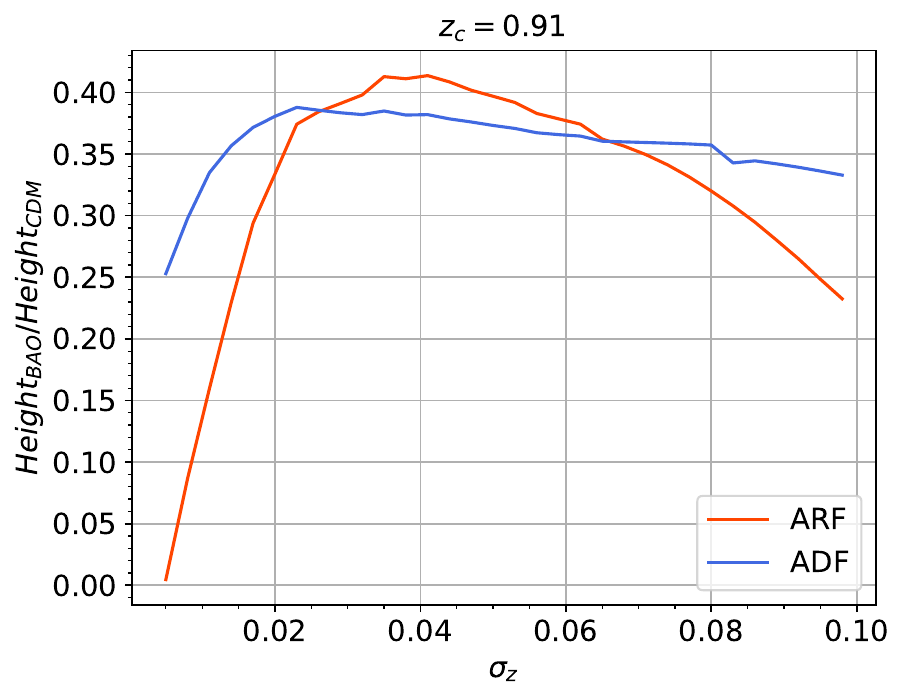}
        \label{fig:rel_heights}
    \end{subfigure}
    \caption{Comparing the relative height of the BAO peak over the zero-lag ($\theta=0$) peak in the auto-correlation function of shells centred at $z_c=0.61$ (left panel) and $z_c=0.91$ (right panel). In both cases, the ratio for the ARF cases (red lines) show a dramatic drop for $\sigma_z \ll (\Delta z)^{BAO}$ ($ (\Delta z)^{BAO}\sim 0.045-0.055$ at those redshifts).} 
    \label{fig:combined_heights}
\end{figure}

This is precisely the behaviour that we recover numerically, shown in Fig.~\ref{fig:combined_heights}. At small values of $\sigma_z$, the ratio of the heights of the BAO peak versus the zero-lag peak decreases notably for ARF (red lines). 
For ADF this suppression is not so steep (blue lines), since according to our interpretation it is only the zero-lag peak that increases its amplitude for low $\sigma_z$s, while the BAO peak, in this case, remains at similar amplitudes. This behaviour remains roughly the same for shell mean redshifts $z_c=0.61$ (left panel) and $z_c=0.91$ (right panel).

\subsection{BAO in the angle and redshift space \texorpdfstring{$\{\theta,\Delta z\}$}{Lg}}\label{sec:bao_red}

Our next step consists of combining shells at different (but nearby) redshifts in order to sample the BAO scale along both redshift and angle axes. For that purpose we compute the angular cross-correlation function between shell pairs ($i,i'$), given in terms of the cross angular power spectrum $C_{\ell}^{i,i'}$ via the standard formula
\begin{equation}
w(\Delta z,\theta) = \sum_{\ell}\frac{2\ell+1}{4\pi} C_{\ell}^{i,i'} P_{\ell} (\cos \theta ).
\label{eq:cross_acf}
\end{equation}
This cross-correlation function can be obtained for either ADF or ARF and for shells lying at different redshift offsets $\Delta z$. The BAO scale should show as a distinct feature whose location should be at ($\theta=\theta_{BAO},\Delta z=0)$ and $(\theta=0,\Delta z=(\Delta z)^{BAO})$ as limiting cases. 

Intuitively one would expect some level of similarity between the 2D angular correlation function $w(\Delta z, \theta)$ and the 3D spatial correlation function $\xi(r)$, whose dependence in $r$ can be further broken into two variables, namely $\xi(r_{\parallel},r_{\perp})$. Due to the redshift space distortions (RSD), the BAO $r_s$ scale does not appear completely spherical in $\xi(r_{\parallel},r_{\perp})$, but slightly distorted due to the radial peculiar velocities that make the BAO sphere to flatten along the radial directions. Interestingly, the impact of radial peculiar velocities in $w(\Delta z, \theta)$ correlation function manifests in a very different way. As shown in, e.g., \cite{hernandez2021density}, radial peculiar velocities contribute to the (auto-)angular power spectrum with a higher amplitude for narrower redshift shells. Since redshift shells are necessarily narrow if we aim to resolve the BAO feature (see discussion in the previous section), the impact of peculiar velocities will be relevant for all shell combinations, and particularly more relevant for lower values of $\sigma_z$. The contribution to the $C_\ell$s from peculiar radial velocities extends to angular scales that are comparable to $\theta_{BAO}$, and thus they add on top of the BAO signature. 

\begin{figure}[htpb]
    \centering
    \includegraphics[width=\textwidth]{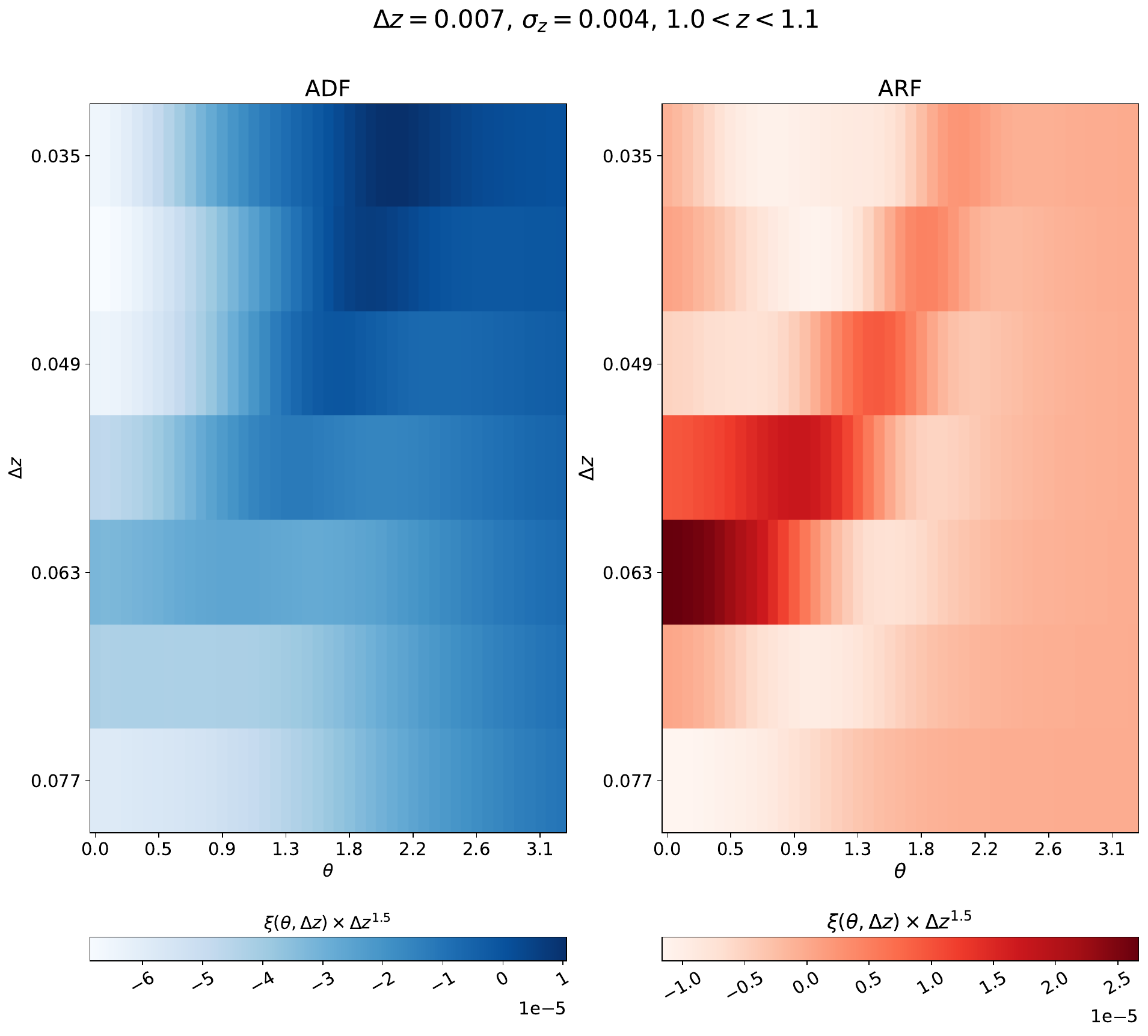}
    \caption{Comparison of two-dimension correlation functions $\xi(\Delta z, \theta) \times \Delta z^{1.5}$ in ($\theta, \Delta z$) space, for ARF (left panel) and ADF (right panel). }
    \label{fig:combined}
\end{figure}

This is shown in Fig.~\ref{fig:combined}, where we are displaying a normalised, 2D angular correlation function defined as
\begin{equation}
    \xi(\Delta z,\theta)\equiv \frac{w(\Delta z,\theta)}{w(\Delta z=0,\theta=0)},
    \label{eq:norm_2d_acf}
\end{equation}
 where the normalised function $\xi(\Delta z,\theta)$ is computed for both ADF (left panel) and ARF (right panel). 
 The darkest colours forming a circular pattern, where the peak at $\Delta z \simeq 0.06, \theta=0$ matches the expected BAO peak position for a $z_c=1.0$ according to Eq.~(\ref{eq:comoving}). While the ADF seem to show a weaker BAO signature for distant redshift bins, this can be improved for narrower shells (see Fig.~\ref{fig:arf_radial_low_z} in the Appendix \ref{ap:ap1}). We see here that, for ARF, narrow shells lose their sensitivity to the BAO peak for the auto-angular spectra, but retain it for cross-angular spectra. Extremely narrow shell configurations, however, are seen problematic even for relatively dense spectroscopic surveys like DESI or {\it Euclid}, since they would be populated with few sources In Appendix~\ref{ap:ap1} we provide an explicit computation of the signal-to-noise ratio of the angular power spectrum under such narrow shells, finding that indeed the $C_\ell$s are just around the shot noise level for {\it Euclid}- and DESI-like surveys. It also turns out that non-linear effects are foreseen to be more important for narrower shells as well, \cite[see, e.g., the comparison of ADF and ARF linear theory prediction to COLA mocks in][]{hernandez2021density}. 

Figure~\ref{fig:combined} displays a weighted version of Eq.~\ref{eq:norm_2d_acf}, i.e, $(\Delta z)^{1.5} \times \xi(\Delta z,\theta) $. This plot corresponds to the {\it Euclid}-like configuration in which we cross-correlate a redshift shell centered on $z_c=1$ with other shells separated by multiples of $\Delta z =0.007$ up to $z=1.1$, under a constant Gaussian width of $\sigma_z=0.004$. In this particular plot the relatively narrow choice of $\Delta z =0.007$ is motivated by the BAO constraint with BOSS data in \cite{marra2019first}. Furthermore, Table~\ref{tab:i} contains the redshift error of the surveys, and for this case (\textit{Euclid}) $\sigma_z =0.004$ corresponds to twice the expected redshift error $\sigma_{\rm Err}$, thus introducing corrections of the order of $\sim 10~\%$ in the effective redshift shell width. As mentioned above (and in Appendix~\ref{ap:ap1} in more detail) it is also shot noise (or finite galaxy number density) on top of redshift errors what compromises the BAO detection under such narrow redshift shells. The BAO scale in the $\{\Delta z, \theta\}$ space seems to show a complementary pattern for ADF and ARF: for the former, it can be more clearly seen at low values of $\Delta z$, while for ARF it seems stronger than for ADF at $\theta \approx 0$. We thus expect that the combination of both observables should improve the sensitivity to the BAO feature. However, its amplitude (given in the colour bars) is relatively weak in both cases. For ADF the BAO peak shows up like a positive peak in a negative background, while for ARF the background is close to zero. 

This approach of conducting multiple shell angular cross-correlations is different from that in \cite{salazar2014clustering}, where the angular auto-correlation function of several redshift shells was rescaled according to an expectation of the angular diameter distance ($D_A(z)$), and binned altogether to increase the signal-to-noise ratio of the BAO feature. That work followed an implementation very similar to that provided in \cite{sanchez2011tracing}. Our study generalizes the work presented in \cite{sanchez2013precise,marra2019first}, where narrow shells were combined to isolate the BAO feature along the redshift axis, but no systematic neighbour shell cross-correlation were considered.

\begin{figure}[htpb]
    \centering
    \includegraphics[width=.8\textwidth]{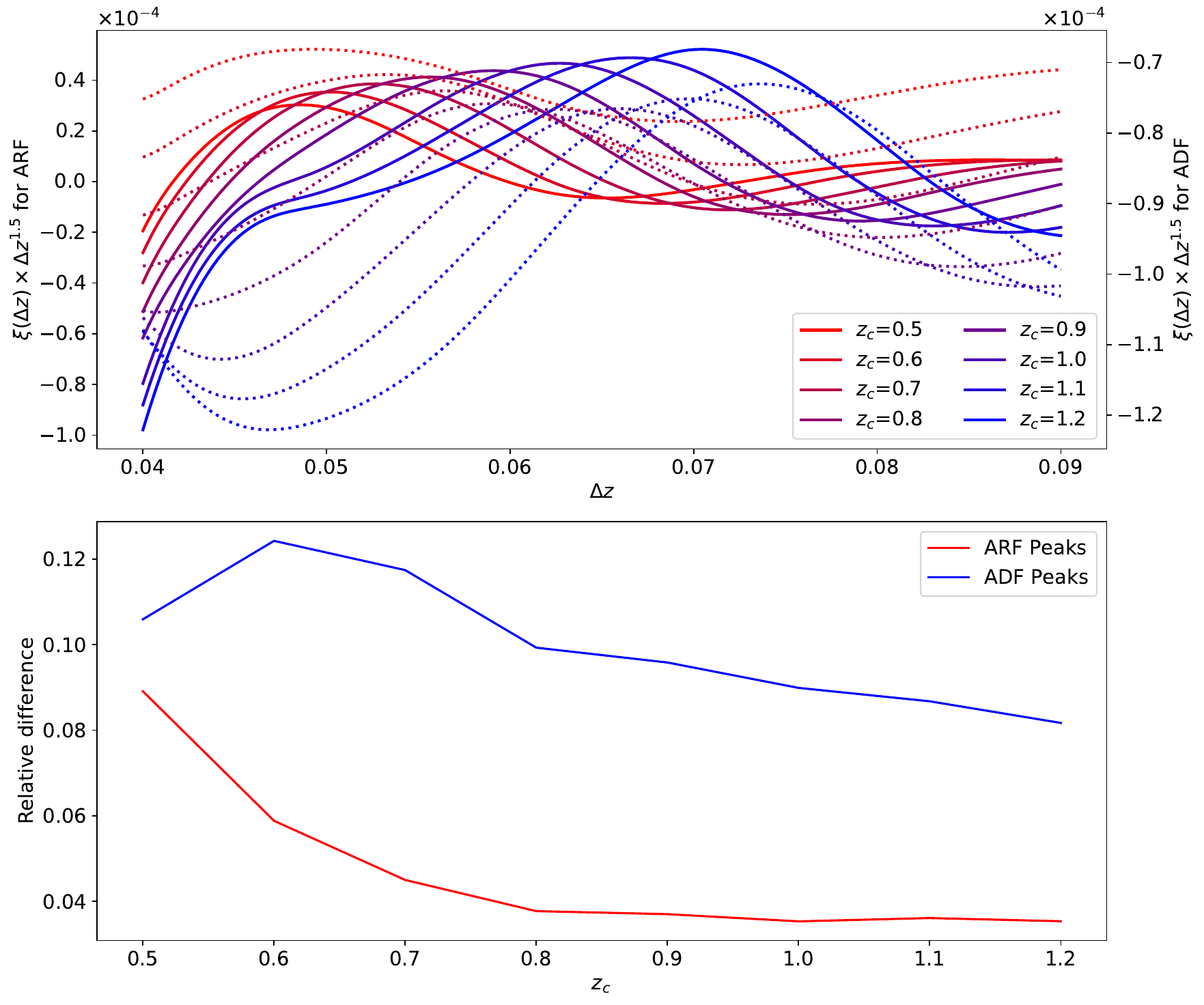}
    \caption{\textit{Upper panel:} Weighted correlation function in redshift space $\xi(\Delta z) \times \Delta z^{1.5}$ for 
 a shell separation of $\Delta z = 0.007$ and a shell width of $\sigma_z=0.004$ for different $z_c$s. Dotted (solid) lines refer to ADF (ARF). \textit{Lower panel}: The relative difference between the BAO peak position from Eq.~(\ref{eq:comoving}) and the position obtained from ADF (blue) and ARF(red) correlation functions.}
    \label{fig:splined_xi}
\end{figure}

In the top panel of Fig.~\ref{fig:splined_xi}, we show the BAO feature in the weighted correlation function along the redshift shift axis $(\Delta z)$ for different values of central redshifts $z_c$, and for ADF (dotted lines) and ARF (solid lines). It follows the expected trend predicted by Eq.~(\ref{eq:comoving}), where the peak position $(\Delta z)^{BAO}$ increases as $z_c$ increases. However, the location of the BAO maxima does not exactly coincide with Eq.~(\ref{eq:comoving}): the bottom panel of this figure shows that, although the ARF seem to agree better with the BAO position from Eq.~(\ref{eq:comoving}) than ADF, both show some difference at the $\sim 12\%$ level. This mismatch was already pointed out at the level of linear theory by \cite[][]{sanchez2008best}, who noted that the BAO scale did not exactly coincide with the position of the BAO peak as predicted by Boltzmann codes like \texttt{CAMB} \cite{lewis2000efficient}, \texttt{CMBFAST} \cite{seljak1996line}, or \texttt{EH98}\cite{eisenstein1999power}. This is not an issue as long as our models capture these shifts properly.

 We are considering relatively narrow shells ($\sigma_z\sim 4\times 10^{-3}<10^{-2}$), and this may prompt some issues from mainly two different sides. The most obvious is the measured redshift precision: it was shown in \citep{hernandez2025} that errors in redshift estimates add in quadrature with the adopted redshift shell width $\sigma_z$, such that the effective width of the redshift window adopted is then (assuming Gaussian errors in redshifts) $\sigma_z^{\rm Eff} = \sqrt{\sigma_z^2 + \sigma_{\rm Err}^2}$, with $\sigma_{\rm Err}$ the error in the redshift measurement. For typical spectroscopic surveys like e.g. BOSS or DESI, $\sigma_{\rm Err}$ remains below the $(1+z) \times 10^{-3}$ level in most cases, and errors as large as $\sigma_{\rm Err}\sim 10^{-3}$ would introduce corrections on the effective redshift width of $\sim 3~\%$ for an initial redshift width of $\sigma_z=0.004$. At the same time, narrow shells are more prompt to non-linear corrections that wider shells, something we have tried to model in this work via the approach of \cite{mead2021hmcode}. Finally, depending on the particular ARF estimator adopted in each work\footnote{As shown in \cite{hernandez2025} there are two different approaches for estimating ARF. One of them behaves robustly for sparse galaxy surveys but is impacted by systematics biasing the observed number of galaxies. The second ARF estimator is much less to those systematics, but becomes numerically unstable for sparse galaxy surveys}, systematics impacting the observed number of galaxies (airmass, star density, extinction, etc) may or may not bias ARF estimates, see \cite{hernandez2025} for further details on the impact systematics on ADF and ARF maps.

\section{Fisher formalism}\label{sec:fisher}

In this section, we implement a Fisher formalism to quantify the amount of new information brought by ARF on top of ADF. We shall consider three different cases, ADF alone, ARF alone, and the combination of both, just as in \cite{legrand2021high}, although in this case we shall quantify the amount of information added by the cross-correlation to a number of neighbouring shells. We will also try to isolate the information associated to the BAO feature by comparing to models with very low $\Omega_b$ (and thus very weak BAO features). In this case of very low baryon content, the amount of dark matter is increased so that the total matter density is identical to our fiducial model. Finally, we will study how the Fisher information depends upon the shell configuration (number of shells versus their intrinsic width $\sigma_z$ for a fixed redshift range).

 We apply this Fisher formalism in both the {\it Euclid}-like and the DESI-like configurations. We work under the assumption of Gaussian statistics and that different multipole $\ell$s are independent from each other (just as in \cite{blanchard2020euclid}), so that Fisher matrices from different $\ell$s can be added. For each $\ell$ we consider any shell of index $c$ that corresponds to a triplet $c \equiv \{o,\beta,\nu\}$, with $o$ referring to the observable (ADF or ARF),  $\beta$ to the redshift bin, and $\nu$ to the neighbour index ($\nu=0$ denotes auto-power spectrum, $\nu\neq 0$ denotes a cross-correlation with a neighbour shell of index $\beta+\nu$). If $c'\equiv \{o',\beta',\nu'\}$, then we rewrite Eq.~\ref{eq:covM0} as
\begin{equation}\label{eq:cov_mat}
    \mathrm{CovM}[C_{\ell}^{c,d},C_{\ell}^{c',d'}] = \frac{C_{\ell}^{c,c'}C_{\ell}^{d,d'}+C_{\ell}^{c,d'}C_{\ell}^{d,c'}}{f_{\rm sky}(2\ell+1)}.
\end{equation}
As in Eq.~\ref{eq:covM0} above, the factor $f_{\rm sky}$ refers to the effective fraction of the sky covered by the survey's footprint. The covariance matrix presents as many blocks as observables are considered. Each block is almost diagonal, with most of the correlation structure along the diagonal and nearby shells.

The data vector for each multipole $\ell$ constitutes the sequence of all angular power spectra considering all observables, redshift shells, and neighbour shells per redshift shell without repetition (running on indexes $c$ and $d$):
\begin{equation}
    D_{\ell}=\{ C_{\ell}^{c,d}\}_{c,d}.
\end{equation}
Finally, the Fisher matrix is written as follows:
\begin{equation}\label{eq:fisher_tensor}
    F_{\alpha \gamma}=\sum_\ell \left(\frac{\partial  D_{\ell}}{\partial \alpha}\right)(\mathrm{\mathbf{CovD}})^{-1}\left(\frac{\partial D_{\ell}}{\partial \gamma}\right),
\end{equation}
where derivatives are computed with a finite difference approach (whose steps are given in Table~\ref{tab:i}) and $\mathrm{\mathbf{CovD}}$ refers to the covariance tensor for the data vector $D_{\ell}$. 

One must note, however, that there are numerical stability issues in the computation of the Fisher matrix, particularly when dealing with the two different observables and multiple neighbours. The fact that the power spectra of the different observables, namely ADF and ARF, have significant magnitude differences, leads to instabilities in the covariance matrix, resulting in negative or null determinants when inverting the matrix. These instabilities are exacerbated by the increasing matrix dimensions if more parameters enter the definition of the Fisher matrix. To address this, two measures are taken. First we renormalize the ARF power spectra by re-scaling the ARF definition by the inverse of $\sigma_z$ (with $\sigma_z$ the width of the Gaussian shells under consideration). This renormalisation should leave the ARF at a similar amplitude level as the ADF. Next we identify and eliminate eigenvalues/eigenvectors in the covariance matrix that are associated with numerical instabilities, by either showing negative or very low signal-to-noise ratio eigenvalues. 
The final Fisher matrix is constructed using valid eigenvalues, and stability is assessed using the Figure of Merit (FoM), which is computed as the determinant of the Fisher matrix of the parameters under study. The results show a clear stability hierarchy across different configurations, with the least noisy being ADF auto-spectra, and the noisiest being the two observable combination (ADF+ARF) with neighbouring shells. We provide further details on the computation of the total Fisher matrix in the Appendix \ref{ap:ni}.

\subsection{Shot-noise}

The two surveys of choice have different redshift distributions and associated (finite) galaxy number densities, which translate into different shot noise levels, and these must be accounted for in our forecasts for the auto power spectra (we neglect shot noise contributions for all cross-power spectra). We follow \cite{legrand2021high} at implementing the expressions for the shot-noise. For the ADF auto angular power spectra, the shot noise contribution is the usual inverse of the number density of galaxies under any given redshift shell $i$ (see Eq.~\ref{eq:ang_no_density}):
\begin{equation}\label{eq:shot_noise1}
    \mathcal{N}^{ADF}_i=\frac{1}{N^i_g},
\end{equation}
where $N^i_g=n^i/4\pi f_{sky}$ ($N_g=\sum_iN^i_g$) is the density of galaxies per unit solid angle on a sphere and $n^i$ is the number of galaxies under a particular shell.
For ARF, the shot-noise expression is $\mathcal{N}^i_{ARF}=1/N^{ARF}_i$, where $N^{ARF}_i$ is:
\begin{equation}\label{eq:shot_noise2}
     N^{ARF}_i=\frac{1}{(N^i_g)^2}\int dz N_g  W^i(z;\sigma_z) \times (z-z_c)^2.
\end{equation}

Contrary to what is done in \cite{matthewson2022redshift}, we neglect the small non-zero shot noise contribution arising in ADF $\times$ ARF cross-correlations since our redshift shells are narrow and that contribution typically appears for wide shells.

When comparing the BAO with the no-BAO cases, we choose to ignore the shot-noise contribution to the covariance matrix. The reason for this is that, in such case, we try to understand how errors in cosmological parameters compare when considering/neglecting BAOs. It turns out that the presence/absence of BAOs change the amplitude of the ADF/ARF angular power spectra, and that errors in parameters are only independent of the $C_\ell$ amplitude if the shot-noise part in the covariance matrix is neglected (since in that case the relative error of the $C_\ell$s is independent of the actual $C_\ell$ amplitude). For instance, for an auto-ADF covariance matrix, we have that:

\begin{equation}
\langle ( C_\ell^{i,i}) ^2  \rangle - \langle C_\ell^{i,i} \rangle^2 = \delta^K_{\ell,\ell'}
    \frac{ 2(C_\ell^{i,i} + \frac{1}{N^i_g} )^2 }{(2\ell+1)f_{\rm sky}},
\label{eq:covM0b}
\end{equation}

where the term $1/N^i_g$ corresponds to the shot noise term. If this term is neglected, the relative error of the $C_\ell$s (and thus the error of the cosmological parameters) are independent of the $C_\ell$ amplitude, enabling a fair comparison of the BAO/no-BAO cases).

\section{Results}\label{sec:results}

\subsection{The error contours}

\begin{figure}[htpb]
    \centering
    \includegraphics[width=1\textwidth]{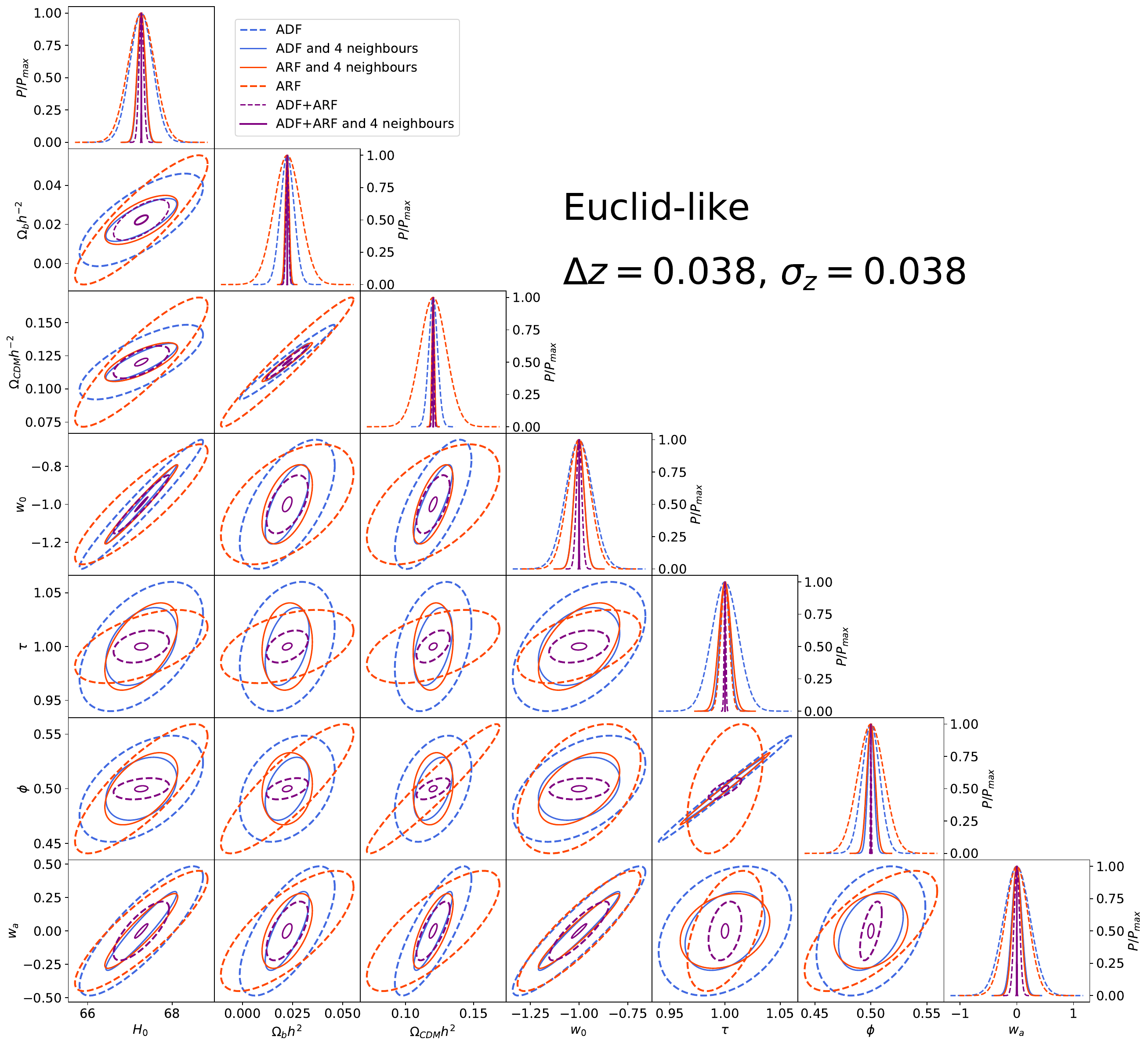}
    \caption{Constraints for a \textit{Euclid}-like survey with shell separation of $\Delta z =0.038$ and shell width of $\sigma_z=0.038$. The dashed lines consider only auto-power spectra in the data vector, while solid lines include the first neighbour of each shell. Red colour displays ARF constraints, blue is used for ADF constraints, and purple for the combination of both observables (ADF$+$ARF).}
    \label{fig:euclid1}
\end{figure}

\begin{figure}[htpb]
    \centering
    \includegraphics[width=1\textwidth]{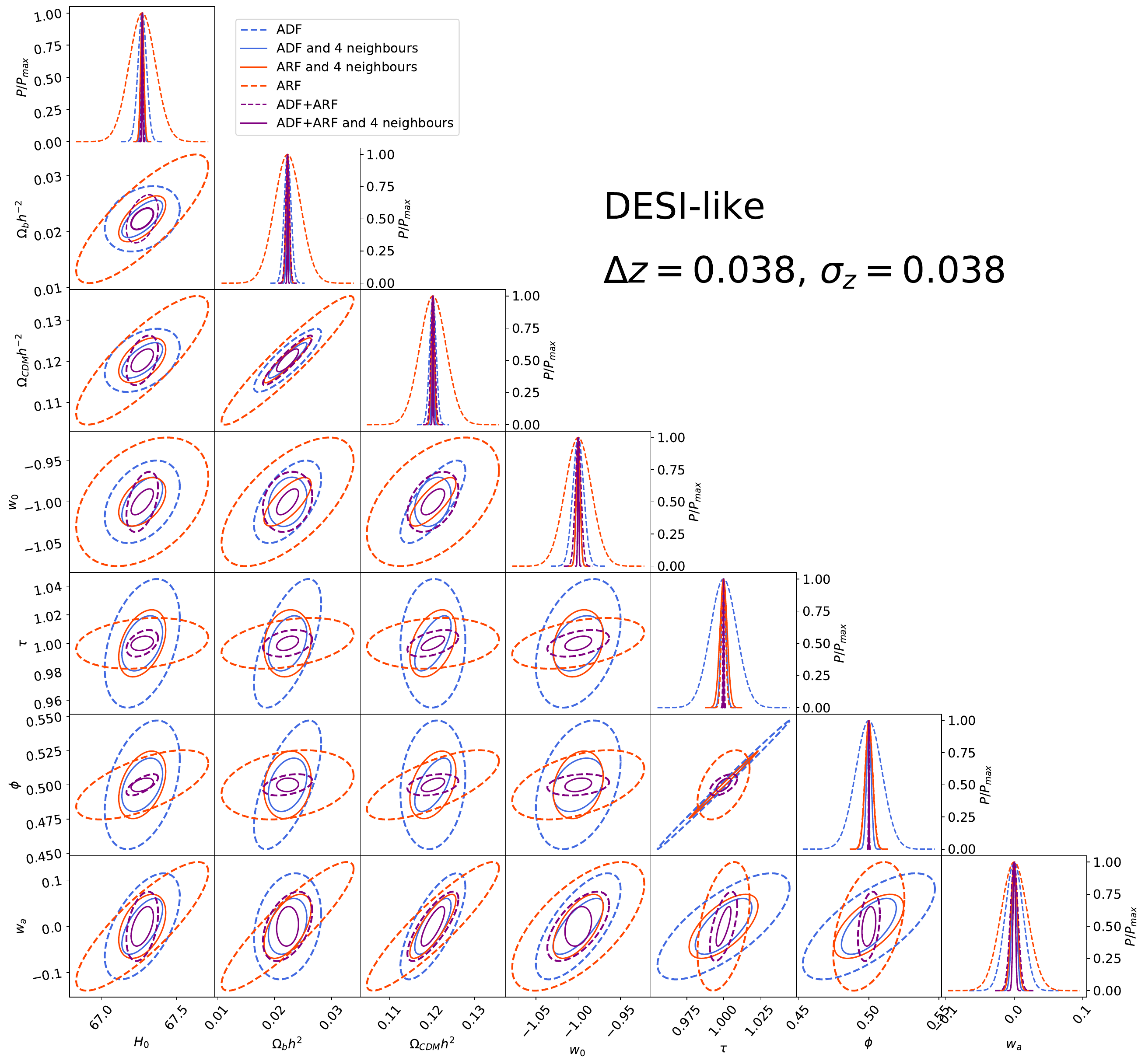}
    \caption{Constraints for a DESI-like survey with shell separation of $\Delta z =0.038$ and shell width of $\sigma_z=0.038$. The dashed lines contain only the auto-power spectra information, while solid lines represent the constraints when the covariance matrix includes the first neighbour of each shell. Orange-red for ARF constraints, blue for ADF constraints, and purple for the Cross-power spectra between observables.}
    \label{fig:desi1}
\end{figure}

We next study the outcome of our Fisher matrix analyses on both {\it Euclid}-like and DESI like surveys. The shell configuration is the default described above, with redshift separation of $\Delta z=0.038$, and equal width of $\sigma_z=0.038$. We first show the error ellipses, first for the {\it Euclid}-like survey (Fig.~\ref{fig:euclid1}) and for the DESI-like (Fig.~\ref{fig:desi1}). As the legend shows, blue, red, and purple colours refer to ADF, ARF, and ADF$+$ARF combined, respectively. Dashed lines refer to the case where we do not consider in our data vector any cross-correlation between different shells. Instead, solid lines display the case when the cross-correlation to four neighbours (the two closest shells), two at higher redshift and two at lower redshift than the centre $z_c$, either for ADF, ARF, or both at the same time. For the sake of simplicity, we do not include in our data vectors cross-angular power spectra involving neighbours with different probes, i.e., the cross-angular power spectra of ADF in one shell and ARF in any of its neighbours. 

The default redshift shell configuration in this work differs from the one used in \cite{legrand2021high} in the choice of $\sigma_z$ ($\sigma_z=0.038$ in this case, $\sigma_z=0.01$ in that work), since here we attempt to maximise the ARF sensitivity to the BAO features. As a consequence, the relative scaling of error contours between ADF and ARF will not exactly follow that one found in \cite{legrand2021high}. In particular, here we find that ARF are more sensitive than ADF to bias-related parameters, while ADF seem to provide slightly more information on the other cosmological parameters given in Table~\ref{tab:cosmo_params}. Nevertheless, and most importantly, the combination of both observables (ADF+ARF) yields significantly more precision to the analysis, in agreement with the findings of \cite{legrand2021high}. 

This is quantified in the error ellipses provided in Figs.~\ref{fig:euclid1} and \ref{fig:desi1} for the {\it Euclid-} and DESI-like configurations, respectively. We find that the orientation of the ADF (blue) and ARF (orange) error ellipses are in practically all cases different enough so that the combination of both probes (magenta) yields a notable improvement. Interestingly, adding the cross-correlation to four neighbour shells (solid lines) shrinks by a factor of a few the area of the error ellipses with respect to the case of no cross-correlation to neighbour shells (dashed lines). This latter result prompts the question on how many neighbour shells should be included in the data vector for Fisher information to converge. This issue is  addressed in Sect~\ref{sec:nu_dep}. 

\begin{figure}
    \centering
    \includegraphics[width=1\textwidth]{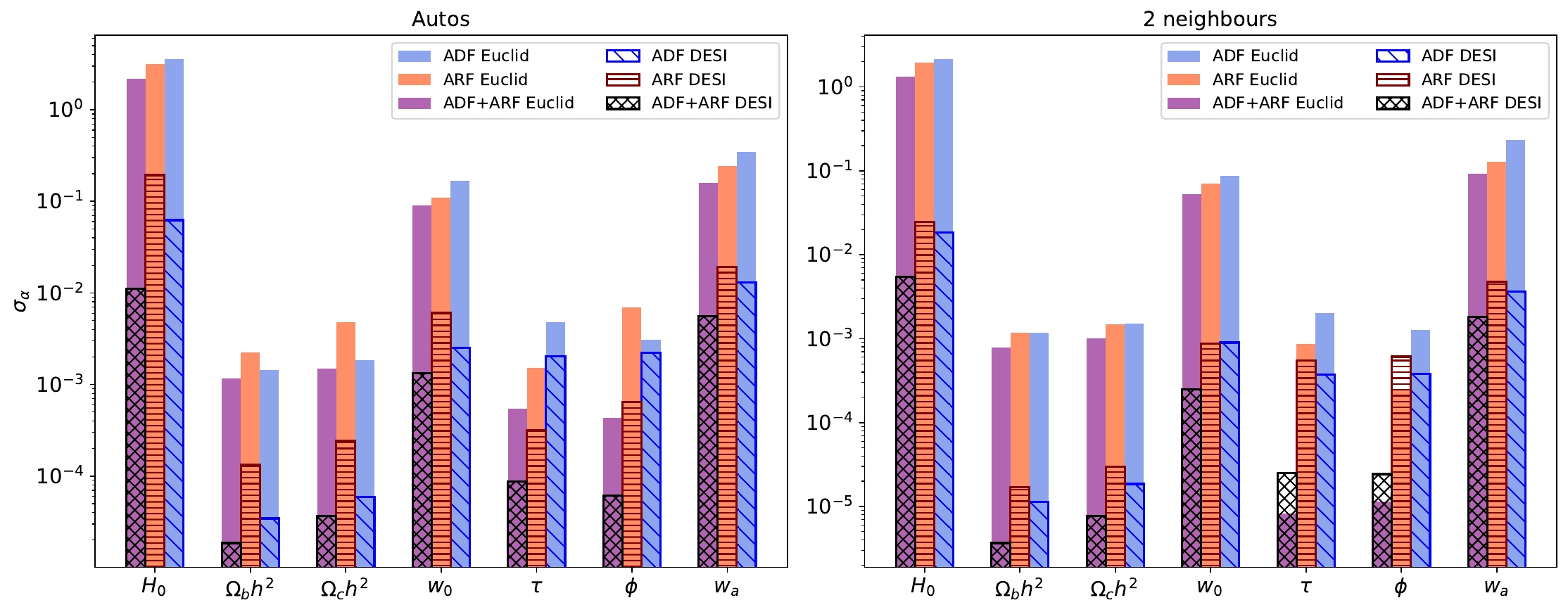}
    \caption{Errors ($\sigma_\alpha$) for each parameter. From left to right the bars show the logarithmic value of the marginalised forecast errors for ADF+ARF, ARF, and ADF. The filled bars represent the \textit{Euclid} survey and the hashed bar, the DESI ones.}
    \label{fig:error_bars}
\end{figure}

We study more in detail the behaviour of the predicted Fisher errors on the parameters in Fig.~\ref{fig:error_bars}: the left panel observes only auto power spectra, i.e., no cross-correlation to any neighbour redshift shell, while the right panel refers to the case where cross-correlation to the two most nearby shells are included in the data vector. For the particular modelisation of the {\it Euclid}-like and DESI-like galaxy surveys, we find that the latter provide smaller uncertainties in all cosmological parameters, and this is presumably due to the higher number density and the lower redshift range sampled by this survey (during which dark energy is more dominant over matter). But overall we can see that the uncertainties in cosmological parameters are comparable from either ADF or ARF, and that those are systematically improved under the ADF+ARF probe combination. This improvement is more dramatic in the bias-related parameters $\tau$ and $\phi$, but relevant in all parameters under consideration. Given the ideal character of Fisher forecasts, further, more realistic work using simulated galaxy mocks will be required to quantify the improvement in the precision on all those cosmological parameters, and in particular those related to physical matter densities and the Dark Energy equation of state. 

 A rough comparison of our error contours with those presented in the Euclid forecasts of \cite{blanchard2020euclid} or in the DESI DR2 results of \cite{karim2025desi} show that, for the ADF+ARF configuration with neighbors, our uncertainties, like theirs, lie in the $1-10~\%$ ballpark, with typically smaller relative uncertainties in  $H_0, \Omega_c, \Omega_c h^2$ than for $\Omega_b, \Omega_b h^2$. Our simplistic description of the two surveys prevents us from conducting more detailed comparison, not being this the scope of this paper.

As in previous works, we are considering a set of cosmological and bias parameters ($\Omega_m h^2$, $\Omega_b h^2$, $A_s$, $w_0$, $w_a$, etc) whose constraints are obtained via a Fisher matrix formalism. In practice, however, analyses have so far focused to extract {\em effective} parameter combinations like $b_g\sigma_8(z), \, Ef\sigma_8(z)$, or even the photometric redshift error $\sigma_{\rm Err}$, \citep{hernandez2021boss,hernandez2025}. Since the bias parameters, in those works, were assumed to be different for each redshift shell, the resulting total number of parameters was relatively high. In future analyses it makes more sense to adopt the strategy used in this work where the bias parameter for all redshift shells is described by a few ($2-4$) entries, so that more constraining power can be dedicated to individual cosmological parameters, much more in line with our Fisher forecasts in this work. We expect natural degeneracies among different parameters combinations, like $\Omega_c h^2$ and $\Omega_b h^2$, or $A_s$ and the bias-related parameters, something that the combination of multiple redshift shells --both in auto- and cross-correlation-- should help to mitigate. This is part of what we explore in this work.

\subsection{The information encoded in the BAO}

This work focuses on the study of the BAO from the ADF and ARF perspectives. Since the Fisher analysis above is generic in the sense it does not particularise on the BAO features, we next try to isolate the amount of information encoded in the BAO wiggles. For that, we consider another cosmological model with identical energy content to our fiducial model so far, but with almost no baryons. We adopt in this case $\Omega_b h^2=0.001$, but maintaining the initial value of $\Omega_m=0.321$. The resulting angular power spectra in this model will effectively show no BAO features, and thus the information obtained from them can be assigned to the shape of the smooth $C_\ell$s. For this analysis, we shall exclude $\Omega_b h^2$ and $\Omega_c h^2$ from the parameter set under study since their fiducial values vary considerably in the BAO/no-BAO scenarios.

\begin{figure}
    \centering
    \begin{subfigure}[b]{0.48\linewidth}
        \includegraphics[width=\linewidth]{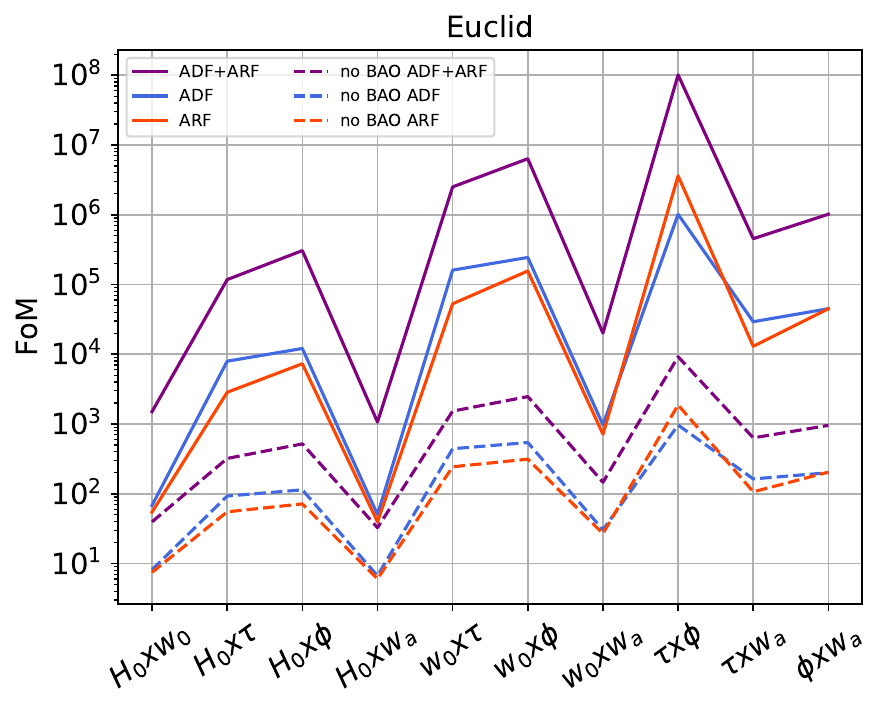}
        \caption{\textit{Euclid} FoM results for some parameters.}
        \label{fig:eu_nobao}
    \end{subfigure}
    \hfill
    \begin{subfigure}[b]{0.48\linewidth}
        \includegraphics[width=\linewidth]{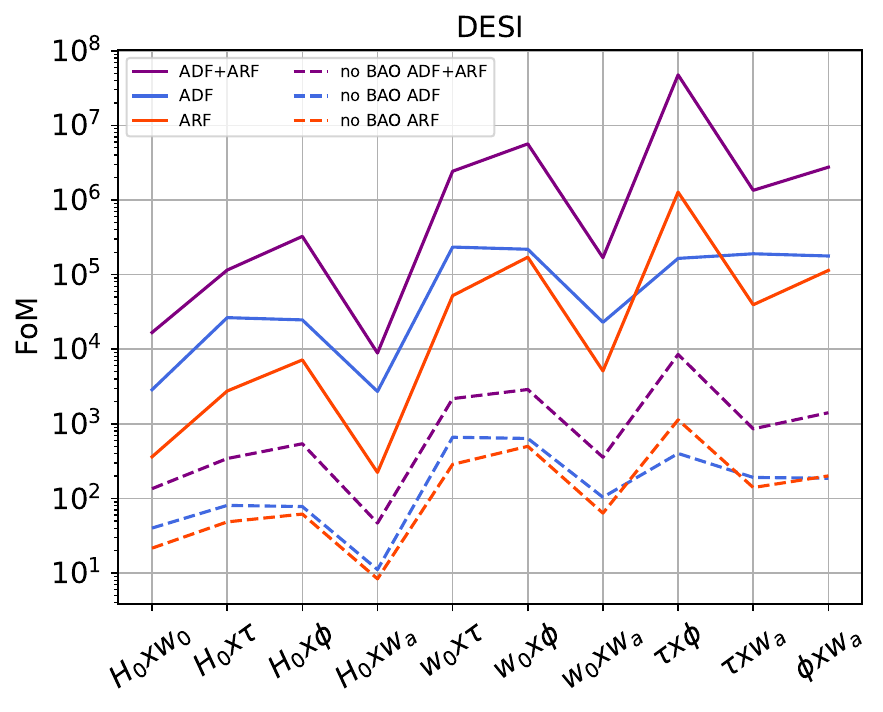}
        \caption{DESI FoM results for some parameter-spaces.}
        \label{fig:desi_nobao}
    \end{subfigure}
    \caption{FoM results for some parameter spaces with (solid line) and without (dashed-line) the BAO feature. The colour scheme is the same as in previous plots.}
    \label{fig:combined_nobao}
\end{figure}

The panels in Fig.~\ref{fig:combined_nobao} display the comparison of the FoMs in the two (BAO/no-BAO) scenarios of the auto power spectra. The solid (dashed) lines correspond BAO (no-BAO) scenarios, for the {\it Euclid}- (left panel) and DESI-like (right panel) surveys.
The behaviour is quite uniform in all cases: ADF and ARF provide similar constraints on the different parameter pairs, but the probe combination of ADF+ARF improve by about an order of magnitude the FoMs for practically all parameter pairs. 

Quite remarkably, the addition of BAO also boosts the FoM in all cases considered, such that the BAO seem to dominate the information content for both experimental configurations and all parameter combinations. This leads to what we interpret as one of the main results of this work: the BAO information from ARF, added on top of that from ADF, should improve by about one order of magnitude the FoM of all the possible parameter combinations, including those involving the equation of state of Dark Energy or the Hubble constant. 

\subsection{Exploring the dependence on neighbour shells \texorpdfstring{$\nu$}{Lg} and shell width (\texorpdfstring{$\sigma_z$}{Lg})}

\label{sec:nu_dep}

So far we have provided forecasts for ADF, ARF, and ADF+ARF for a fixed shell configuration that targets the sensitivity to the BAO features. In this subsection, instead, we explore the dependence of the Fisher forecasts on different shell configurations, namely on the number of neighbouring shells, and the shell width $\sigma_z$ adopted.
In what follows, we consider always an even number of neighbour shells per ``central" shell, i.e., one at lower and one at higher redshift\footnote{Obviously, for the extreme shell centred upon the lowest and highest redshift considered only one neighbour will be possible, but yet the total number of neighbours will be even.}.

\begin{figure}[htpb]
    \centering
    \begin{subfigure}[b]{.48\textwidth}
        \centering
        \includegraphics[width=\textwidth]{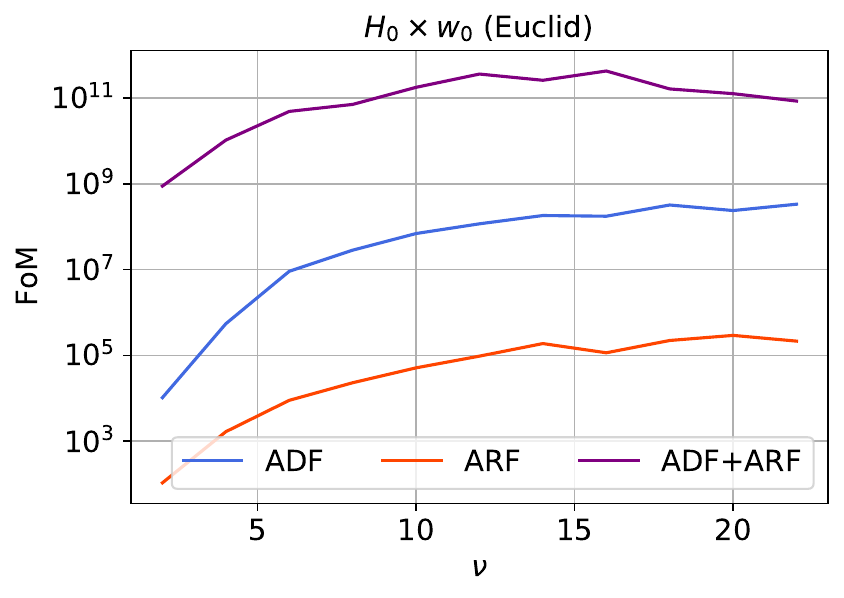}
        \caption{}
        \label{fig:H0w0}
    \end{subfigure}
    \hfill
    \begin{subfigure}[b]{.48\textwidth}
        \centering
        \includegraphics[width=\textwidth]{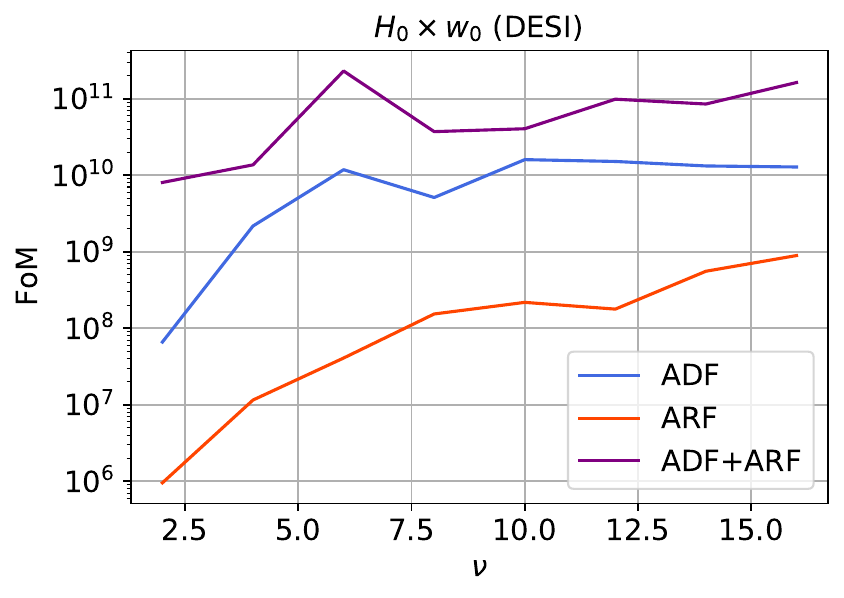}
        \caption{}
        \label{fig:H0w0_desi}
    \end{subfigure}
    
    \caption{FoM for the $H_0-w_0$ parameter space w.r.t. the number of neighbours ($\nu$).}
    \label{fig:h0w0_fom}
\end{figure}

\begin{figure}[htpb]
    \centering
    \begin{subfigure}[b]{.48\textwidth}
        \centering
        \includegraphics[width=\textwidth]{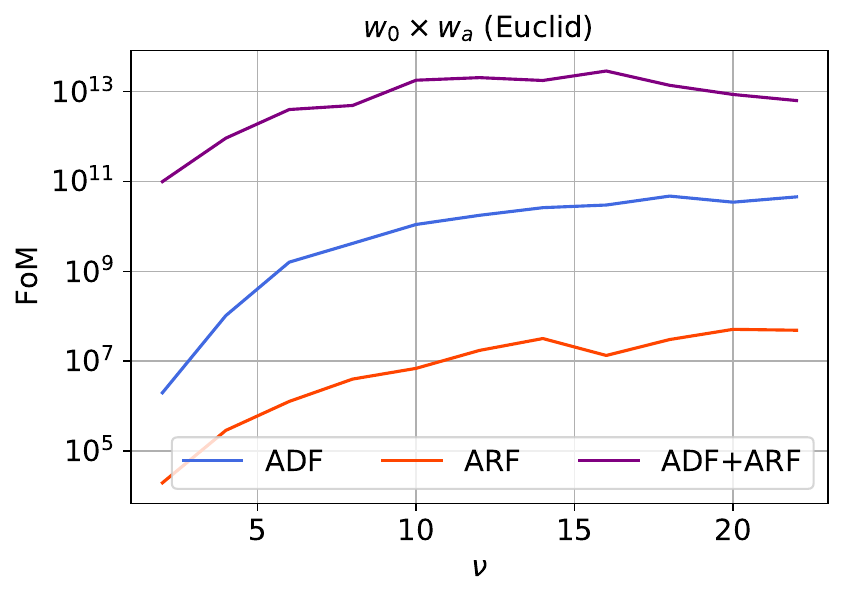}
        \caption{}
        \label{fig:w0waa}
    \end{subfigure}
    \hfill
    \begin{subfigure}[b]{.48\textwidth}
        \centering
        \includegraphics[width=\textwidth]{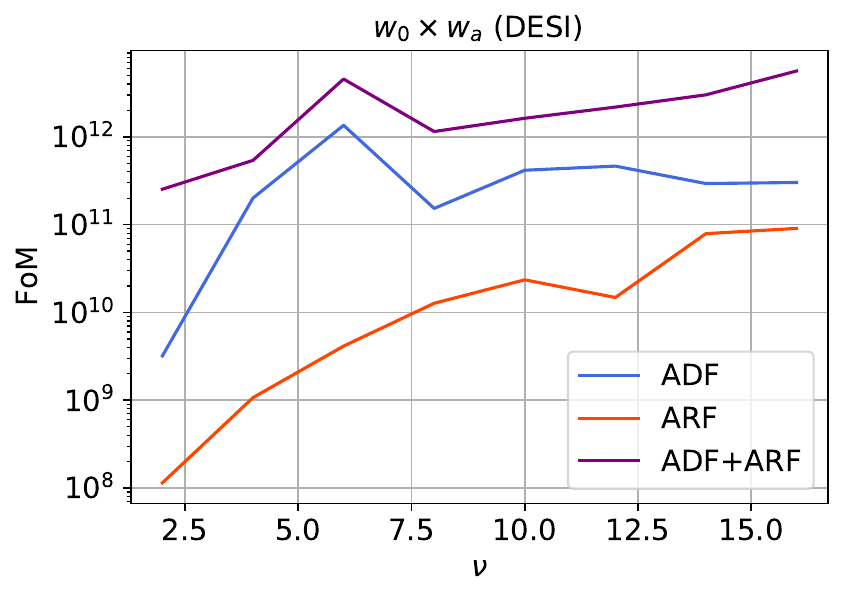}
        \caption{}
        \label{fig:w0wab}
    \end{subfigure}
    
    \caption{FoM for the $w_0 - w_a$ parameter space w.r.t. the number of neighbours ($\nu$).}
    \label{fig:w0wa}
\end{figure}

While naively one would expect that a handful number of neighbour shells would be required to capture the correlation lying between different shells, we actually find a rather different behaviour for ADF and for ARF. 
Figure~\ref{fig:h0w0_fom} displays the FoM for the $H_0 - w_0$ parameters with respect to the number of neighbours ($\nu$), while in Fig.~\ref{fig:w0wa} refers the case for $w_0 - w_a$. These are rather representative cases for all other parameter configurations. We see that while the FoM from ARF saturate at around $\nu\sim 10$ neighbour shells, the ADF FoM instead tends to require more neighbour shells (typically $\nu\sim 15$) to converge. This translates into a higher FoM for ADF, although the FoM due to the combination of ADF and ARF (magenta lines) is always more than an order of magnitude higher than either the ADF-only or the ARF-only cases.

This different behaviour is not surprising for the relatively wide shells of this configuration ($\sigma_z=0.038$). The velocity term in the source terms for both ADF and ARF is significantly smaller than the density term for such choice of $\sigma_z$ \citep{hernandez2021density}, and while the kernel for the density term in ADF allows for large inter-shell correlations, the ARF kernel carries instead the $(z-z_c)$ factor. In practice, this $(z-z_c)$ factor acts as a redshift/radial gradient operator under the redshift shell, making the ARF sensitive to the galaxy density redshift evolution {\rm under} each shell, unlike the ADF, which are sensitive to its average density. Thus the redshift correlation structure for the ARF is based on how radial/redshift gradients correlate from shell to shell, whereas for ADF such correlation is simply built upon the matter/galaxy correlation on large scales.  

The ADF FoM relative contribution to the ADF+ARF FoM results remarkably higher for the DESI-like than for the {\it Euclid}-like configuration, and this seems to be related to the larger growth of the ADF FoM for $\nu \in [5,10]$ in the former case. The only obvious difference between the two experimental configurations is the redshift range they are sampling. The DESI-like is sampling lower redshifts where the transition from matter to dark energy domination is more evident than for the {\it Euclid}-like case. This intrinsic redshift evolution seems to be giving more weight to the cross-correlation with different redshift shell than for the higher redshift range probed by {\it Euclid}.   

\begin{figure}[htpb]
    \centering
    \subfloat[Figure of Merit w.r.t $\sigma_z$ for \textit{Euclid} ($1.0<z<1.2$). \textit{Blue}: ADF. \textit{Red}: ARF. \textit{Purple}: Cross.\label{fig:full_fom_euclid}]{
        \includegraphics[width=.45\textwidth]{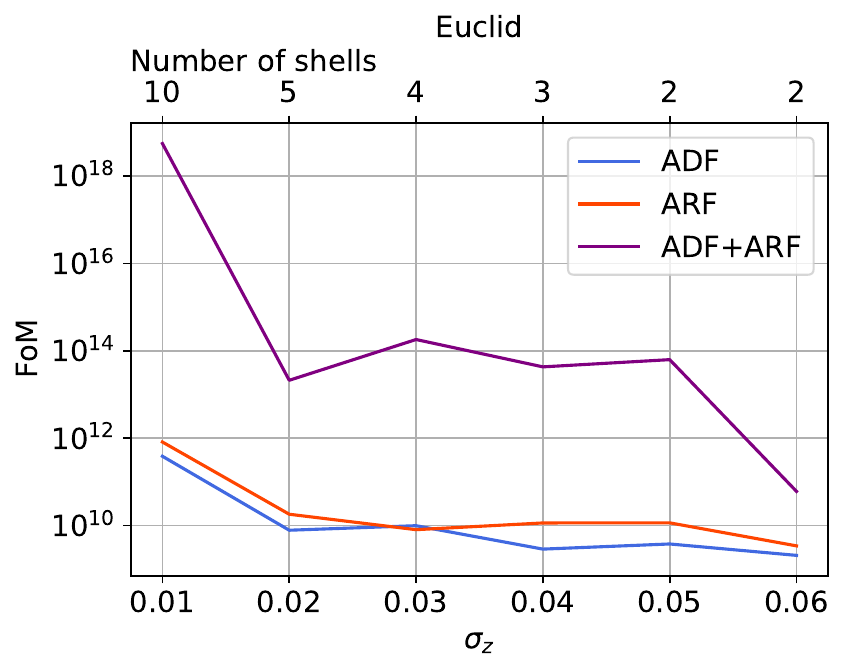}
    }
    \hfill
    \subfloat[Figure of Merit w.r.t $\sigma_z$ for DESI ~($0.6<z<0.8$). \textit{Blue}: ADF. \textit{Red}: ARF. \textit{Purple}: Cross. 
    \label{fig:full_fom_desi}]{
        \includegraphics[width=.45\textwidth]{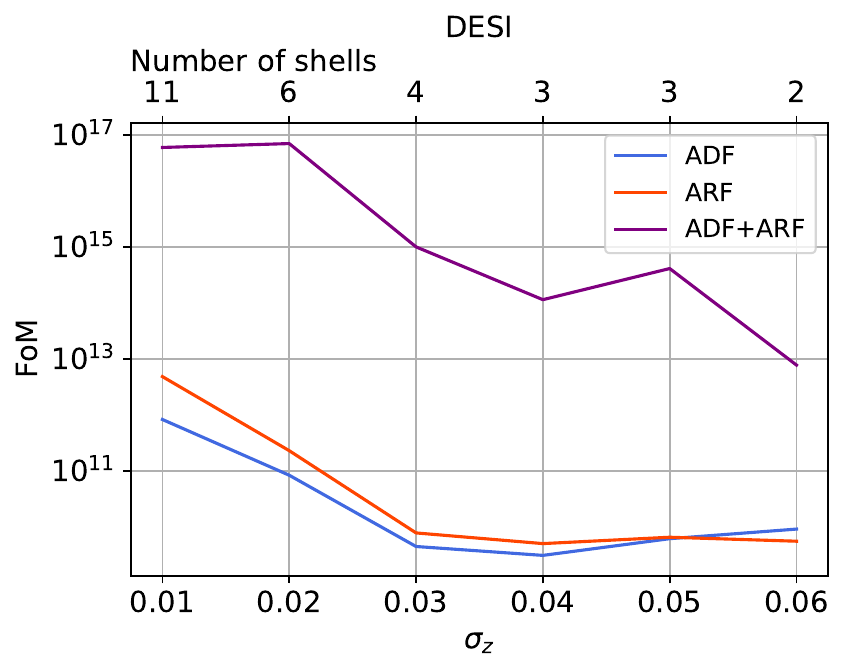}
    }
    \caption{Figure of Merit w.r.t $\sigma_z$ for \textit{Euclid} and DESI.}
    \label{fig:full_fom}
\end{figure}

We now generalize the $\sigma_z=0.038$ shell configuration motivated by the sensitivity to BAO and explore how the FoM behave under different shell configurations. It was shown in \cite{hernandez2021density} that redshift shells contain more cosmological information if they are sufficiently narrow so the term induced by radial peculiar velocities contributes significantly (this term becomes negligible to widths typically larger than $\sigma_z\sim 0.02-0.03$). We have found in previous sections that, while for ARF the ratio of the BAO peak is lost for very narrow shells, it is not the case when cross-correlation between different shells are included in the data vector. So if the effective sensitivity to BAO remains at low values of $\sigma_z$, we would expect the FoM to decrease for increasing $\sigma_z$. When testing this, we fix a redshift interval (from $z=1.0$ up to $z=1.2$ for the {\it Euclid}-like setup, and from $z=0.6$ up to $z=0.8$ for the DESI-like one), and change $\sigma_z$ while fixing the redshift shell separation at $\Delta z = 2\sigma_z$. This results in a varying (decreasing) number of redshift shells as $\sigma_z$ increases. Figure~\ref{fig:full_fom} displays the behaviour of the FoM (built upon {\em all} parameters under study) versus $\sigma_z$ for ADF, ARF, and both probes combined, ADF$+$ARF. The top axis provides the effective number of shells. Different values of $\sigma_z$ may correspond to the same (integer) effective number of redshift shells: in these cases, there will be a different degree of overlap between shells (higher for wider $\sigma_z$).

\begin{figure}[htpb]
    \centering
    \begin{subfigure}[b]{0.49\textwidth}
        \centering
        \includegraphics[width=\textwidth]{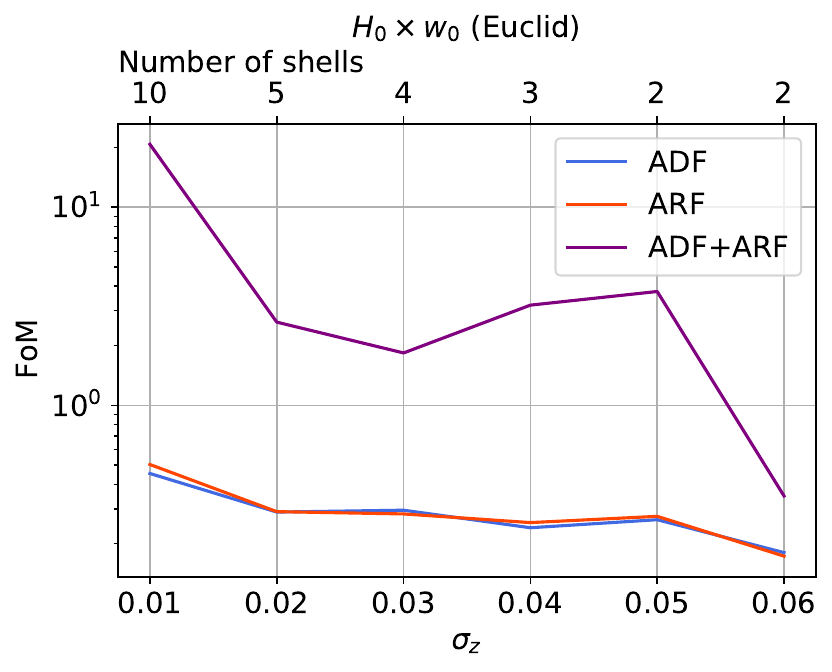}
        \caption{\textit{Euclid} ($1.0<z<1.2$)}
        \label{fig:euclidfom_h0}
    \end{subfigure}
    \begin{subfigure}[b]{0.49\textwidth}
        \centering
        \includegraphics[width=\textwidth]{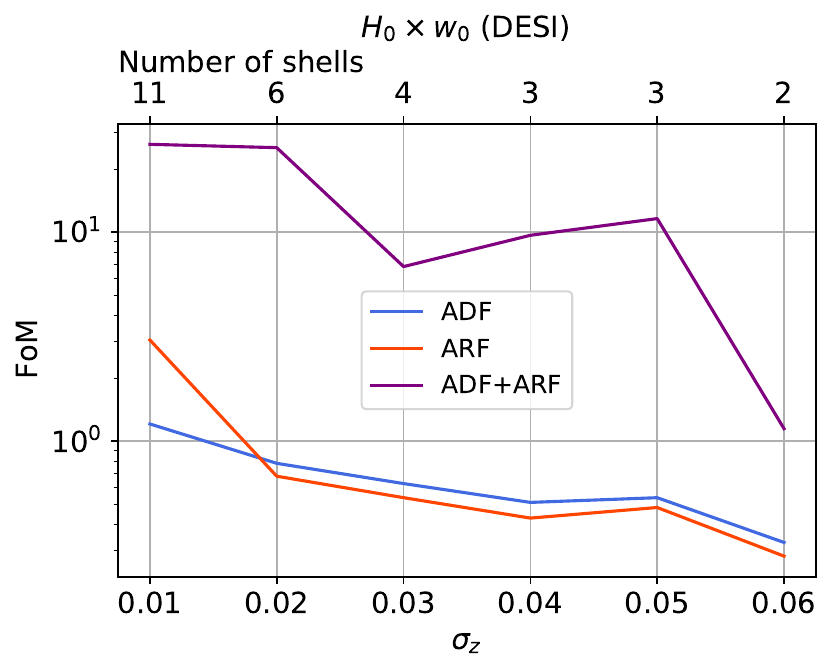}
        \caption{DESI ($0.6<z<0.8$)}
        \label{fig:desifom_h0}
    \end{subfigure}
    \caption{Figure of Merit for $H_0 \times w_0$ w.r.t. $\sigma_z$ for \textit{Euclid} and DESI.}
    \label{fig:fom_h0}
\end{figure}

\begin{figure}[htpb]
    \centering
    \begin{subfigure}[b]{0.49\textwidth}
        \centering
        \includegraphics[width=\textwidth]{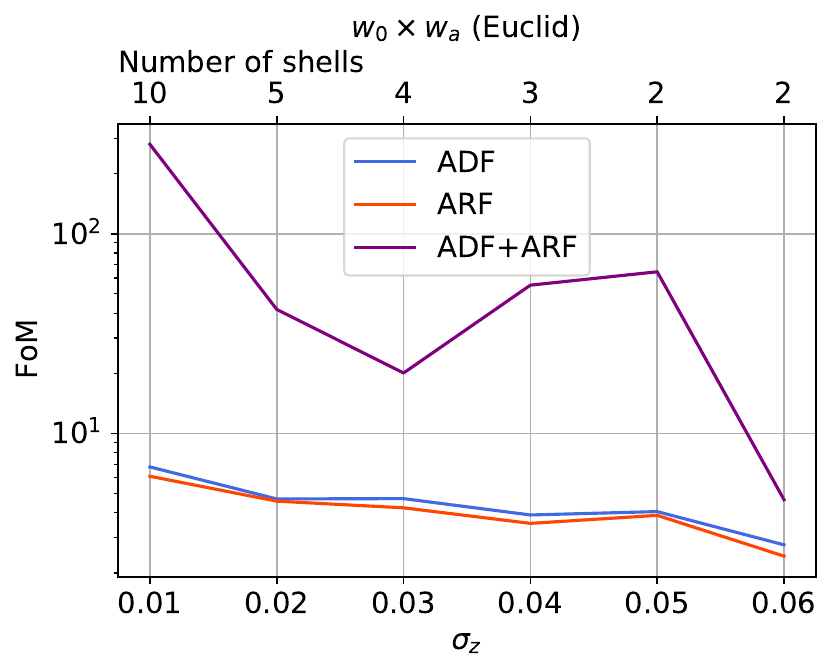}
        \caption{\textit{Euclid} ($1.0<z<1.2$)}      
        \label{fig:euclidfom_w0}
    \end{subfigure}
    \begin{subfigure}[b]{0.49\textwidth}
        \centering
        \includegraphics[width=\textwidth]{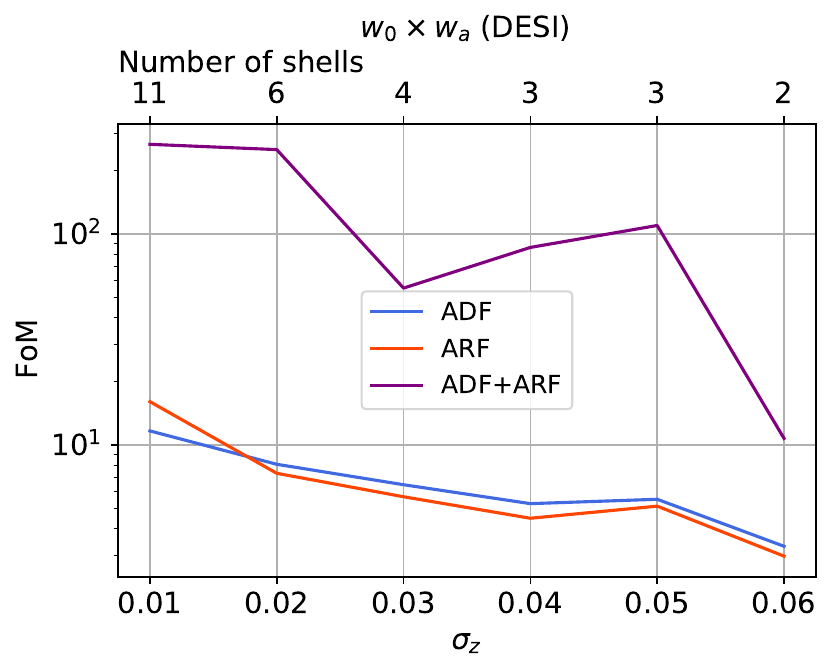}
        \caption{DESI ($0.6<z<0.8$)}
        \label{fig:desifom_w0}
    \end{subfigure}
    \caption{Figure of Merit for $w_0 \times w_a$ w.r.t. $\sigma_z$ for \textit{Euclid} and DESI.}
    \label{fig:fom_w0xwa}
\end{figure}

\begin{figure}[htpb]
    \centering
    \begin{subfigure}[b]{0.49\textwidth}
        \centering
        \includegraphics[width=\textwidth]{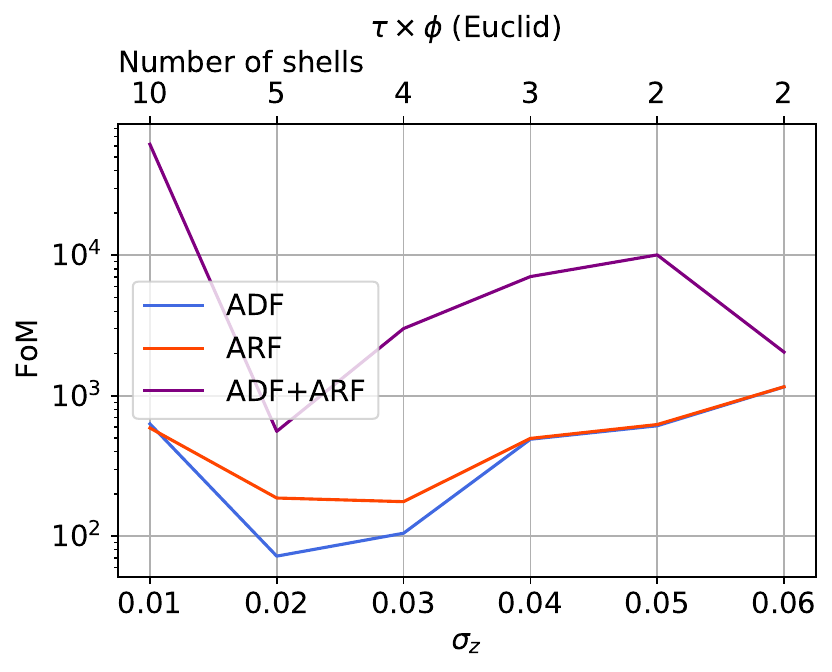}
        \caption{\textit{Euclid} ($1.0<z<1.2$)}      
        \label{fig:euclidfom_bias}
    \end{subfigure}
    \begin{subfigure}[b]{0.49\textwidth}
        \centering
        \includegraphics[width=\textwidth]{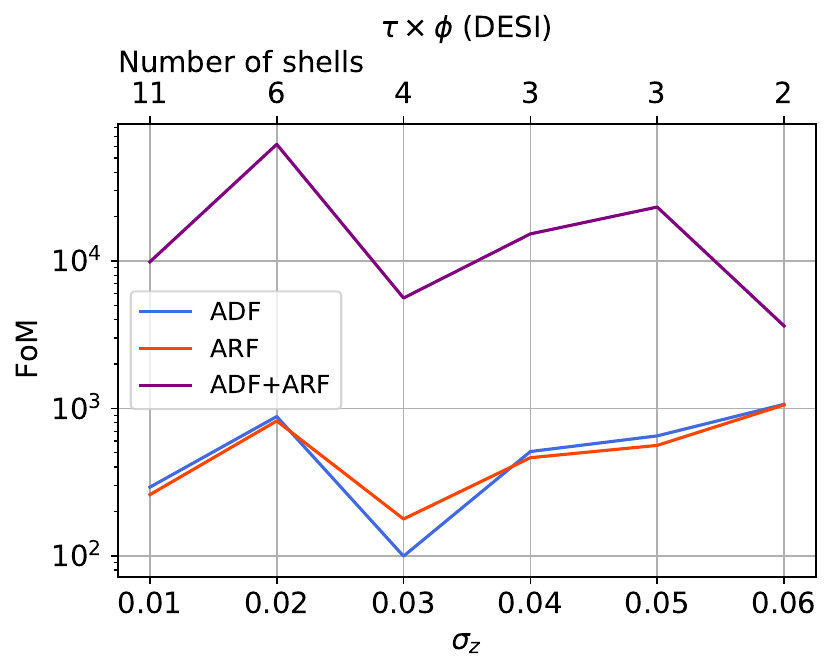}
        \caption{DESI ($0.6<z<0.8$)}
        \label{fig:desifom_bias}
    \end{subfigure}
    \caption{Figure of Merit for $\tau \times \phi$ w.r.t. $\sigma_z$ for \textit{Euclid} and DESI.}
    \label{fig:fom_bias}
\end{figure}

Overall we recover the expected trend: the amount of information decreases steeply when $\sigma_z$ grows above a few times $0.01$, a change that we expect if the velocity term falls well below the density term in the source function $S(k,r)$ for both ADF and ARF. For the highest values of $\sigma_z$ the FoMs enter a plateau that is recovered when we repeat these analyses for the specific parameter combinations $H_0 \times w_0$ (Fig.~\ref{fig:fom_h0}) and $w_0 \times w_a$ (Fig.~\ref{fig:fom_w0xwa}). 
It is interesting that, in Fig.~\ref{fig:full_fom}, the amplitude and shape of the plateau are very similar for the two experimental configurations ({\it Euclid}-like and DESI-like), despite their differences in the redshift ranges sampled by each of them. 
A similar picture can be seen in Fig.~\ref{fig:fom_w0xwa} (for $w_0 \times w_a$) and in Fig.~\ref{fig:fom_bias} (for the bias parameters $\tau \times \phi$). 

We see in Figs.~\ref{fig:fom_h0},~\ref{fig:fom_w0xwa}
similar shapes and amplitudes, pointing to similar sensitivities to these parameter pairs in the $z\in [0.6,0.8]$ and $z\in [1,1.2]$ redshift ranges. They suggest that the inclusion of velocities (and smaller radial perturbation modes that are resolved by the redshift shells) contribute with a significant amount of sensitivity to those cosmological parameters. 

Interestingly, the constraints of the bias-related parameters shown in Fig.~\ref{fig:fom_bias} behave very differently. Such parameters depend upon the typical halo mass mostly and have no direct dependence upon cosmology. Contrary to all the three previous examples, the FoM shows an increasing trend with $\sigma_z$ for ADF-only and ARF-only: even for the same number of redshift shells, the FoM is larger for wider $\sigma_z$s (see for instance $\sigma_z=0.05,\,0.06$, for which the number shells equals $2$). The FoM is higher for $\sigma_z=0.06$, and this can be due to a higher overlap of the two Gaussian windows, which translates into a higher number density and lower effective shot noise level. The interpretation for low $\sigma_z$s is not straightforward either, since both experimental setups show different behaviours. Given the similarity at high $\sigma_z$s in both panels, differences of the FoM at low widths can again be associated with differences in the velocity contribution to the source term in both ADF and ARF. However, in this case, and given the higher number of redshift shells involved in the FoM computation, we do not discard numerical instabilities impacting the FoM shapes, which would also impact the ADF$+$ARF case (we discuss this possibility further in the upcoming section).

\section{Discussion and conclusions}\label{sec:discussion}

According to the Fisher analysis presented in this work, BAO carry most of the information in tomographic analyses of a set of 2D redshift shells extracted from spectroscopic LSS surveys like {\it Euclid} or DESI, either for ADF, ARF, or ADF+ARF combined. We first find that this is the case when only auto-angular power spectra are considered for each redshift shell of width $\sigma_z\sim0.034$. This is a particular configuration for which the BAO peak is well visible in the angular auto-correlation function for both ADF and ARF. ARF carry a similar amount of information as ADF, and the combination of both increases the Fisher matrix determinant by more than an order of magnitude for practically all cosmological parameter pairs under study. This reproduces the results of \cite{legrand2021high}, although in that work the redshift shell configuration observed narrower shells ($\sigma_z=0.01$), and no particular consideration was given by the amount of information carried by the BAO. 

We build further upon all previous results and generalise the shell configuration, including arbitrary widths and cross-correlation to neighbour redshift shells. Overall, the main results hold under all these new configurations: ARF contribute with a significant amount of new constraining power on the cosmological parameters, in particular $H_0$ and the Dark Energy CPL ones ($w_0$,$w_a$), and the combination of both probes ADF and ARF is always superior to either of the probes (ADF or ARF) alone. Including the cross-correlation to a number of neighbour shells increases the constraining power on the cosmological parameters: while the FoM from ARF typically converges for $\nu\simeq 10$ neighbour shells (for $\sigma_z =\Delta z = 0.038)$, resulting in a total redshift range of $(\Delta z)^{\rm tot} \simeq 0.4$), ADF seem to require extending up to a higher number of neighbours ($\nu\sim 15$, or $(\Delta z)^{\rm tot} \simeq 0.6$) in order to find some evidence of FoM flattening versus $\nu$. Due to the $(z-z_c)$ factor present in the ARF kernel, ARF are known to measure radial/redshift gradients {\em under} each redshift shell, and this would make them less sensitive than ADF to the structure under different (and distant) redshift shells. This would also explain why the ADF FoM grows above that of ARF as soon as cross-correlations to neighbour shells are included in the data vector. But even in those cases, the addition of the ARF significantly increases the FoM, probably due to the inclusion of information arising in radial modes stemming from redshift shells that the ADF are blind to\footnote{ADF are sensitive to the monopole/average galaxy number density under the redshift shell, and thus un-sensitive to fluctuations of that number density under it. }.

We further generalize the choice of $\sigma_z$ in our tomographic study. By first time, we recover the BAO peak in the cross-shell angular correlation function in $\{\theta, \Delta z\}$ space for both ADF and ARF, and although weak, the contrast of the BAO peak is actually higher for very narrow shells ($\sigma_z\sim 2-5 \times 10^{-3}$). In this context, we study the constraining power on cosmological parameters given by the determinant of the Fisher matrix under different redshift shell widths for a fixed cosmological volume, such that wider shell configurations correspond to a lower total number of redshift shells. Narrower shells are able to resolve radial shorter wavelength perturbation modes, and thus should access more cosmological information. This is indeed the general trend we find, for all probe combination (ADF, ARF, and ADF+ARF), and for both experimental setups considered ({\it Euclid}- and DESI-like). Among all parameters, only the bias-related ones seem to show a different scaling with $\sigma_z$, with less clear evolution in the range of $\sigma_z$/number of redshift shells sampled: the change in the FoM from the narrowest to the widest shell configuration changes less than a factor of $\sim 2$. Unlike the other cosmological parameters, the linear bias parameters ($\tau \times \phi$) are un-sensitive to the $C_\ell$ shape or the BAO feature (which in general depends upon the shell width $\sigma_z$), but merely measure the amplitude of the $C_\ell$s with respect to the linear prediction. This result suggests that the loss of information caused by the smearing of the radial short-wavelength modes under wider redshift shells is roughly compensated for by a decrease in shot-noise error under such shells.

Our Fisher matrix analyses assume cosmological linear perturbation theory (with the addition of the non-linear corrections of \cite{mead2021hmcode}), Gaussian statistics for the data vectors involved, and lack of correlation between different multipoles of the angular power spectra under study. These are rather strong (and optimistic) assumptions, and real parameter uncertainties are bound to be larger in real data analysis. Even when we have carefully removed noisy eigenmodes possibly biasing our Fisher metrics (FoMs and error forecasts), our predictions are likely to change substantially in a realistic setup. The inclusion of a high number of neighbour redshift shells and derived cross angular power spectra results in large Fisher matrices whose manipulation and interpretation must be done carefully, even if the results we have presented here have proved robust versus changes in the definition of noisy eigenmodes (see Appendix~\ref{ap:ni}). The impact of coupling for different harmonics due to a limited sky coverage, non-linear corrections in both the density and radial peculiar velocity of the tracers, and non-Poissonian shot noise are all aspects that need to be quantified further. They constitute the subject of our forthcoming work on this topic.

\section{Code availability}
The particular software packages used in this work will be accessible at \url{https://github.com/psilvaf/ARF_supplemental.git} at publication. The {\tt ARFCAMB} Boltzmann code is available upon request at \url{https://github.com/chmATiac/ARFCAMB}.

\appendix
\section{Two-dimensional correlation function in redshift space}\label{ap:ap1}
\begin{figure}[htpb]
    \centering
    \includegraphics[width=.8\textwidth]{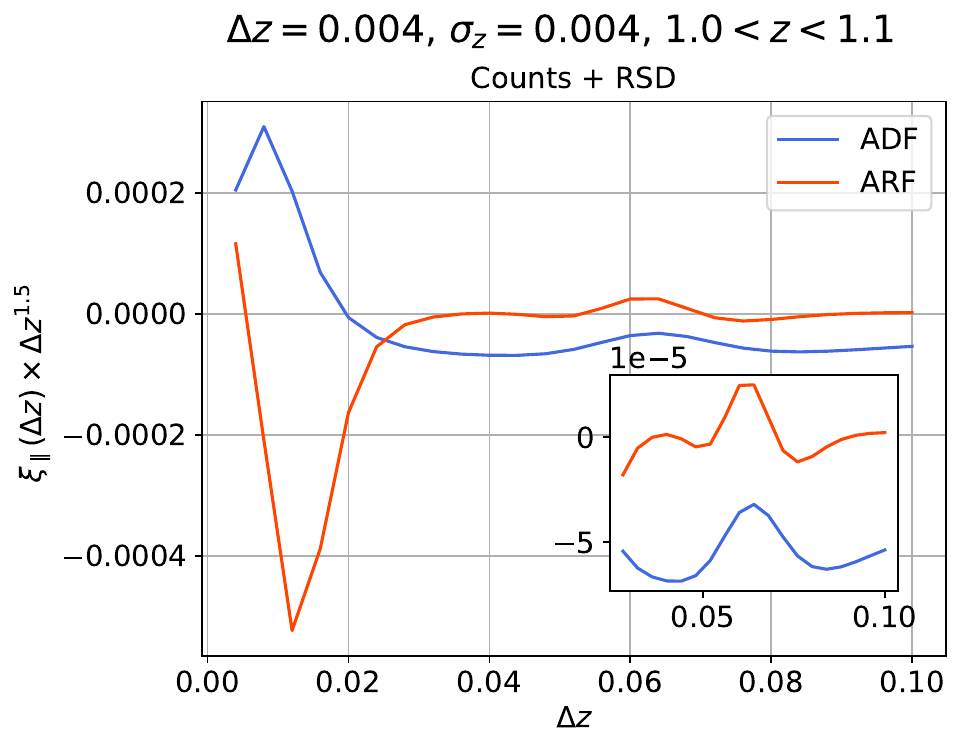}
    \caption{Correlation function in redshift space $\xi_\parallel(\Delta z)$ after including the leading radial peculiar velocity term, the so-called {\tt redshift} term in {\tt ARFCAMB}.  We use a shell separation of $\Delta z = 0.004$ and a shell width of $\sigma_z=0.004$ for a central redshift of $z_c=1.0$ that is cross-correlated to other shells up to $z=1.1$. 
    }
    \label{fig:radial_low_z}
\end{figure}

\begin{figure}[htpb]
    \centering
    \includegraphics[width=.8\textwidth]{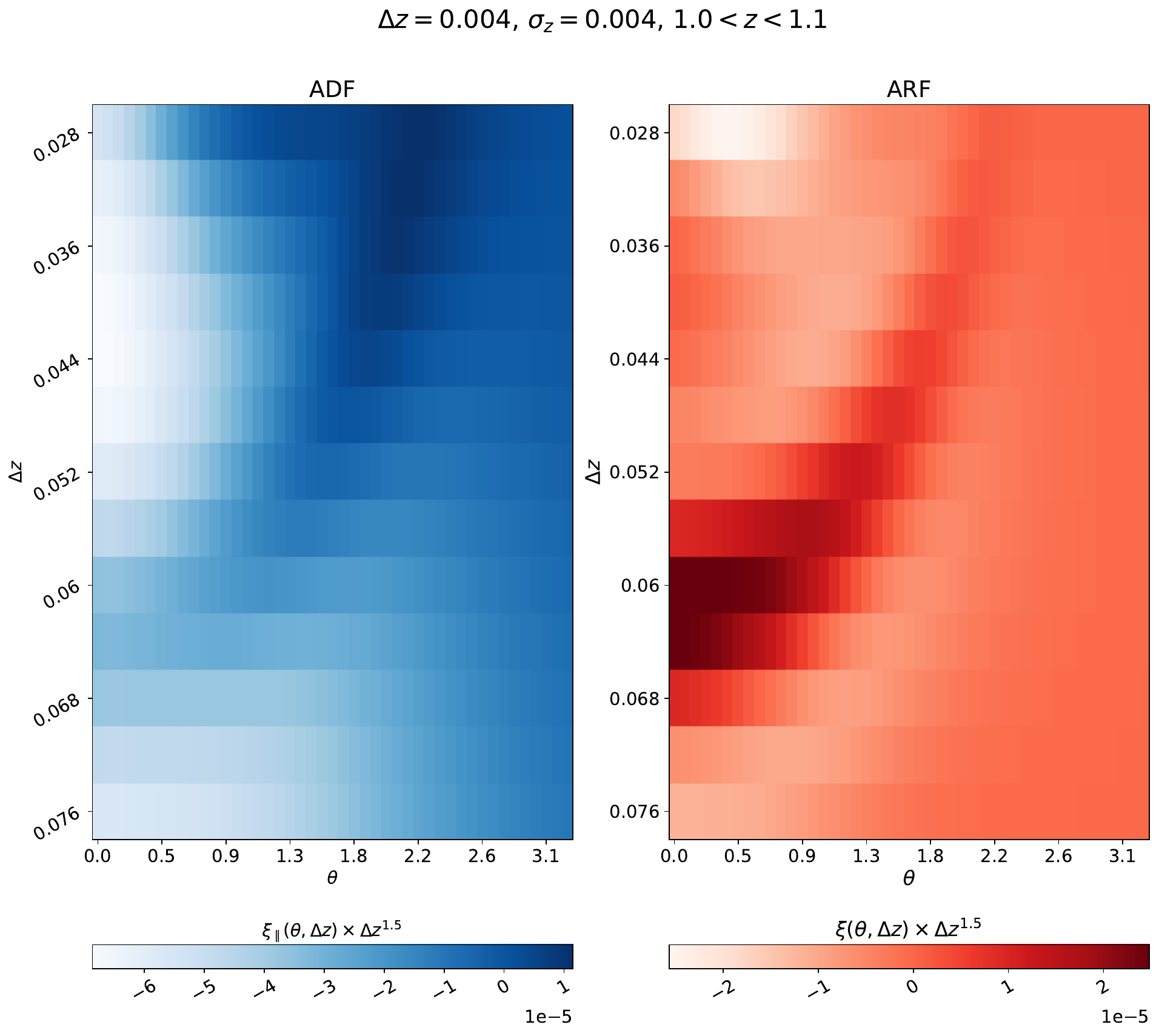}
    \caption{Two-dimensional correlation function $\xi(\Delta z, \theta)$ from ARF in $\{\theta, \Delta z\}$ space for a central shell on $z_c=1.0$ of width $\sigma_z=0.004$ that is cross-correlated to other shells spaced in steps of $\Delta z = 0.004$ up to $z=1.1$.
     }
    \label{fig:arf_radial_low_z}
\end{figure}

We show here another example for the $\{\theta, \Delta z\}$ correlation function of ADF/ARF which considers yet thinner redshift shells that give rise to a clearer BAO pattern. The zero-lag case is displayed in Fig.~\ref{fig:radial_low_z}, while the full, 2D maps is provided in Fig.~\ref{fig:arf_radial_low_z}. In the first case, the inset panel highlights the BAO feature at the expected $\Delta z$ value according to Eq.~(\ref{eq:comoving}), and in agreement with the findings of \cite{marra2019first}. 

\subsection{Signal to noise ratio in the context of the radial BAO from thin shells}

We next check the detectability for narrow choices of $\sigma_z$ using the signal-to-noise ratio coefficient in the following equation:
\begin{equation}
    S/N^2(\ell) = \frac{(C_\ell^{ij})^2}{\sigma^2\left[C_\ell^{ij}\right]},
\end{equation}
where
\begin{equation}
\sigma^2\left[C_\ell^{ij}\right] = \frac{1}{f_{\text{sky}}(2\ell + 1)} \left\{
    \left(C_\ell^{ii} + N_\ell^{ii}\right)\left(C_\ell^{jj} + N_\ell^{jj}\right) + 
    \left(C_\ell^{ij}\right)^2
\right\}.\end{equation}
In order to add the signal coming from all harmonics, we integrate over multipoles as follows:
\begin{equation}
    S/N = \sqrt{\sum_\ell S/N^2(\ell)}.
\end{equation}

\begin{figure}
    \centering
    \includegraphics[width=.8\linewidth]{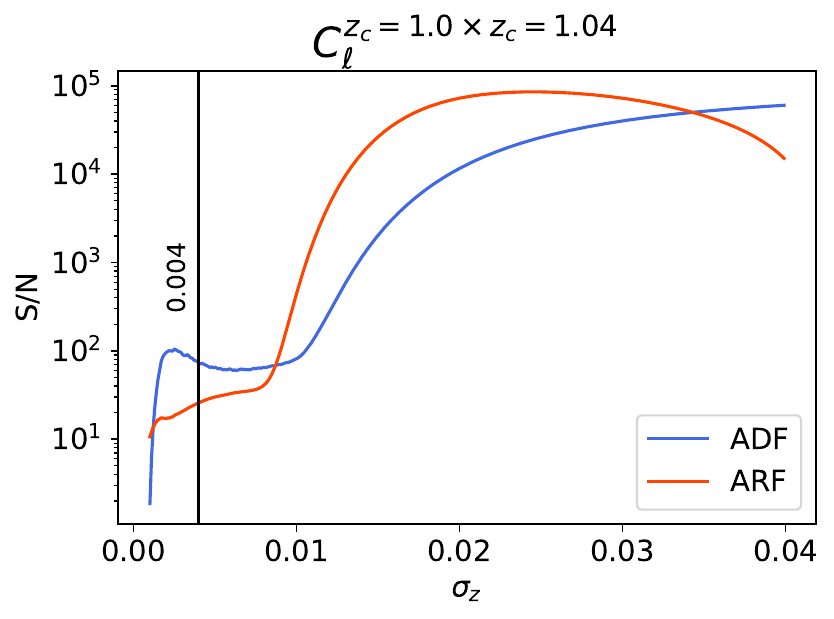}
    \caption{Signal to noise ratio in terms of $\sigma_z$ of ARF (\textit{red}) and ADF (\textit{blue}) when crossing shells with $z_c=1.0$ and $z_c=1.04$, the black vertical line represents $\sigma_z=0.004$, the Gaussian width chosen in $\xi(\Delta z, \theta)$.}
    \label{fig:snr}
\end{figure}

Because we are discussing the radial signal from Eq.~\ref{eq:norm_2d_acf} for $\xi(\Delta z,\theta)$, we consider the cross-correlation of two distinct redshift sehlls, one centered uopon $z_c=1.0$ and another one on $z_c=1.04$, thus separated in by $\Delta z =0.04$. We vary $\sigma_z$ in a range including the adopted value of $\sigma_z=0.004$ in our analysis, and the result is displayed in Fig.~\ref{fig:snr}. For very narrow shells, the signal-to-noise ratio ($S/N$) lies just above the $S/N\sim 10$ level, leaving marginal room for any BAO detection. For wider shells, ARF contains more signal, in agreement with discussions in Sects.~\ref{sec:bao_feat_sep} and \ref{sec:bao_red}.

\section{Numerical stability}\label{ap:ni}

We adopt Eq.~\ref{eq:fisher_tensor} for the Fisher matrix, although here we particularise for each harmonic $\ell$ at a time and simplify its notation as
\begin{equation}
F_{\alpha\gamma} = \biggl(\frac{\partial \mathbf{d}}{\partial \theta_\alpha}\biggr)^t
\mathrm{\mathbf{CovM}^{-1}} \biggl(\frac{\partial \mathbf{d}}{\partial \theta_\gamma}\biggr),
\label{eq:Fish1}
\end{equation}
Here, $\mathbf{d}$ represents the data vector encompassing both auto- and cross-angular power spectra of ADF and/or ARF across any set of redshift shells. The covariance matrix for $\mathbf{d}$ is denoted by $\mathrm{\mathbf{CovM}}$. To ensure a stable inversion of this matrix and an accurate calculation of the Fisher matrix and its determinant, the following procedure is employed:
\begin{enumerate}
\item We standardise the amplitude of all components in our data vector $\mathbf{d}$. The vector may include ADF or ARF angular power spectra, with ARF generally having amplitudes that are $\sigma_z$ times that of ADF. To balance this, we normalise the ARF auto-spectra by the square of the shell width, $(\sigma_z)^2$, and by $\sigma_z$ the cross-spectra involving ARF only once. This adjustment equalises component amplitudes, resulting in the renormalized data vector $\mathbf{\tilde{d}}$ and its related covariance matrix $\mathrm{\mathbf{Cov{\tilde M}}}$.
\item We diagonalize $\mathrm{\mathbf{Cov{\tilde M}}}$, decomposing it as
\begin{equation}
\mathrm{\mathbf{Cov{\tilde M}}} = {\cal R}^t (\Lambda \mathbf{I})\cal{R}, 
\label{eq:diag}
\end{equation}
where $\cal{R}$ is an orthogonal rotation matrix, and $(\Lambda \mathbf{I})$ is a diagonal matrix whose diagonal elements are given by $\lambda_i$, with $i=1,N$, and $N$ the dimension of the data vector $\mathbf{\tilde{d}}$. The columns of the rotation matrix are the eigenvectors $\mathbf{e}_i$ of the renormalized covariance matrix. We discard then all eigenvectors for which $\lambda_i\leq 0$ or $|\mathbf{e}_i\cdot\mathbf{e}^\star_i-1|>\epsilon$\footnote{The symbol $\star$ denotes here ``complex-conjugate".}, with $\epsilon$ an arbitrarily small number to be defined (we adopted $\epsilon\in [10^{-5},10^{-3}]$ in our analyses). Imaginary parts of the eigenvalues are ignored. 
\item Finally, from the remaining pool of eigenvectors, we drop those eigenvectors having a low signal-to-noise ratio. That is, after rotating the data vector 
\begin{equation}
\mathbf{f} = \cal{R}\mathbf{d},
\label{eq:eigenv}
\end{equation}
we further ignore those vectors for which $f_i/\sqrt{\lambda_i}>\eta$, with $\eta$ again an arbitrary small number. 
\end{enumerate}

We make sure that our results are stable with respect to the choices of $\epsilon$ and $\eta$. Our final configuration adopts $\epsilon=10^{-4}$ and $\eta=10^{-2}$: much stricter choices would provide conservative/pessimistic estimates of the FoM, while relaxing those conditions may incur in a significant amount of numerical noise.

We evaluate stability for each multipole with the Figure of Merit (FoM):
\begin{equation}
    \mathrm{FoM}=\sqrt{|F_{\alpha\beta}|}
\end{equation}
calculated as the square root of the determinant of the Fisher matrix from Eq.~(\ref{eq:fisher_tensor}). 

The results in Table~\ref{tab:fom_results_compact} demonstrate that the $\eta$ parameter dominates the performance across all methodologies and multipoles. For ADF at $\ell=70$, relaxing $\eta$ from $10^{-3}$ to $10^{-1}$ causes noticeable performance degradation ($1.059 \rightarrow 0.998$), while ARF at $\ell=20$ shows similar sensitivity with $\eta=10^{-1}$ performing substantially worse ($0.861$). Most dramatically, the combined ADF+ARF method experiences catastrophic failure at $\ell=20$ for $\eta=10^{-1}$, with FoM ratios collapsing to $2.65\times10^{-5}$ and $1.28\times10^{-9}$ for $\nu=0$ and $\nu=10$ respectively. In contrast, optimal performance stabilizes across all configurations when $\eta=10^{-2}$, yielding near-perfect ratios ($\approx 1.000$) regardless of $\epsilon$ variations. This consistent pattern confirms that $\eta$ is the primary parameter controlling numerical stability, while $\epsilon$ plays a secondary role, primarily affecting results only when $\eta$ is suboptimally configured.

\begin{table}[h]
\centering
\caption{FoM$/$FoM$^{fid}$ values in \textit{Euclid-like} configuration for $\ell=10,70$ and $\nu=0,10$, where $fid$ means our default configuration ($\epsilon=10^{-4}$, $\eta=10^{-2}$).}
\label{tab:fom_results_compact}
\begin{tabular}{l l S[table-format=1.3] S[table-format=1.3] S[table-format=1.3] S[table-format=1.3]}
\toprule
\multirow{2}{*}{Method} & \multirow{2}{*}{Parameters} & \multicolumn{2}{c}{$\ell=20$} & \multicolumn{2}{c}{$\ell=70$} \\
\cmidrule(lr){3-4} \cmidrule(lr){5-6}
 & & {$\nu=0$} & {$\nu=10$} & {$\nu=0$} & {$\nu=10$} \\
\midrule
\multirow{4}{*}{ADF}
 & $\epsilon=10^{-4}$, $\eta=10^{-1}$ & 0.956 & 0.999 & 0.998 & 0.999 \\
 & $\epsilon=10^{-4}$, $\eta=10^{-3}$ & 1.003 & 1.000 & 1.059 & 1.000 \\
 & $\epsilon=10^{-5}$, $\eta=10^{-2}$ & 1.000 & 1.000 & 1.000 & 1.000 \\
 & $\epsilon=10^{-3}$, $\eta=10^{-2}$ & 1.000 & 1.000 &  1.000 & 1.000 \\
\addlinespace
\multirow{4}{*}{ARF}
& $\epsilon=10^{-4}$, $\eta=10^{-1}$ & 0.861 & 1.000 & 0.911 & 1.000 \\
 & $\epsilon=10^{-4}$, $\eta=10^{-3}$ & 1.000 & 1.001 & 1.033 & 1.000 \\
 & $\epsilon=10^{-5}$, $\eta=10^{-2}$ & 1.000 & 1.000 & 1.000 & 1.000 \\
 & $\epsilon=10^{-3}$, $\eta=10^{-2}$ & 1.000 & 1.000 & 1.000 & 1.000 \\
\addlinespace
\multirow{4}{*}{ADF+ARF}
& $\epsilon=10^{-4}$, $\eta=10^{-1}$ & \text{2.65e-5} & \text{1.28e-9} & 1.000 & 1.000 \\
 & $\epsilon=10^{-4}$, $\eta=10^{-3}$ & 1.000 & 1.001 & 1.000 & 1.000 \\
 & $\epsilon=10^{-5}$, $\eta=10^{-2}$ & 1.000 & 1.000 & 1.000 & 1.000 \\
 & $\epsilon=10^{-3}$, $\eta=10^{-2}$ & 1.000 & 1.000 & 1.000 & 1.000 \\
\bottomrule
\end{tabular}
\end{table}

\begin{figure}
    \centering
    \begin{subfigure}[b]{.55\linewidth}
        \includegraphics[width=\linewidth]{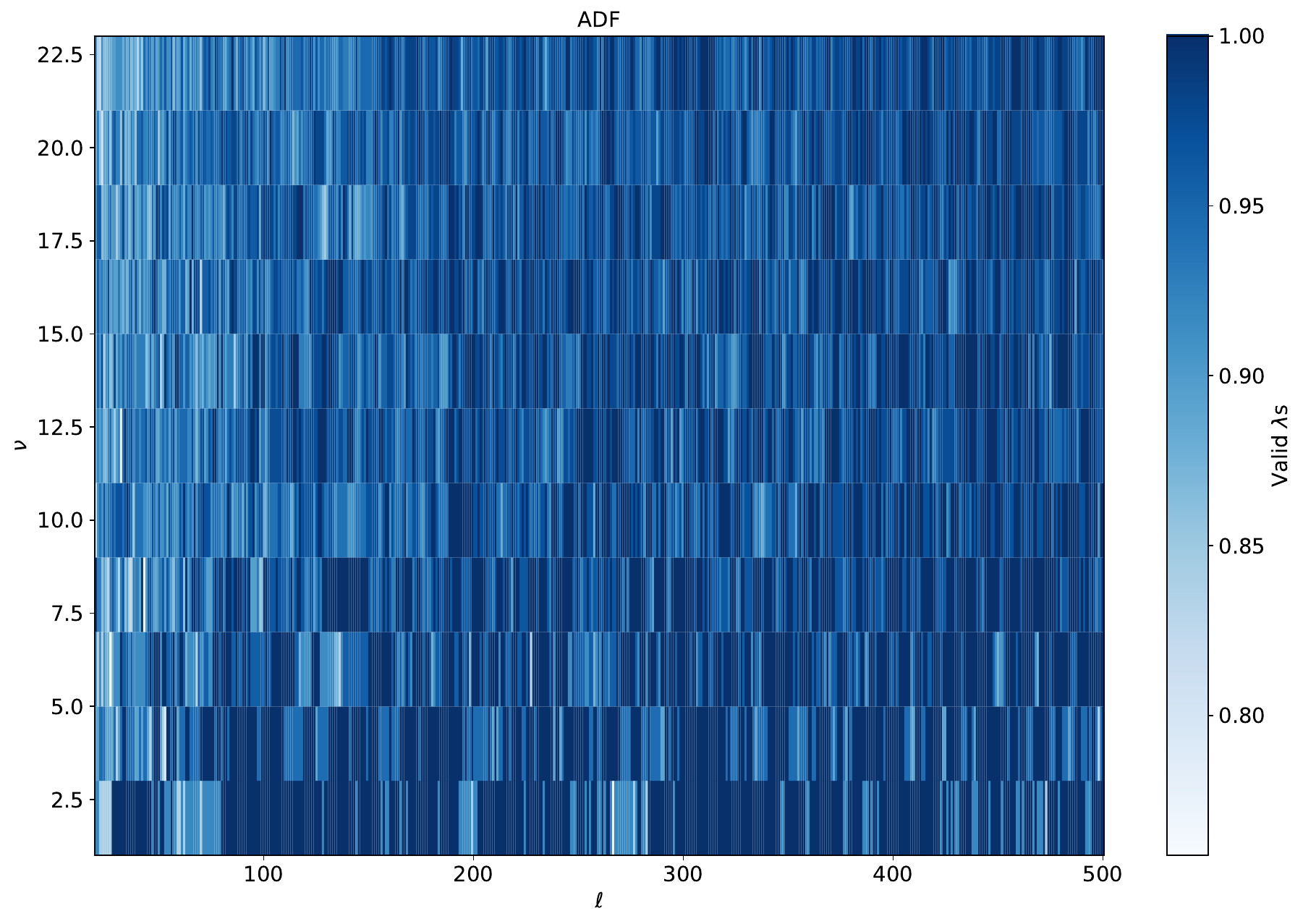}
        \caption{ADF valid eigenvector fraction versus  $\nu$ and $\ell$.}
        \label{fig:ell1}
    \end{subfigure}
    
    \begin{subfigure}[b]{.55\linewidth}
        \includegraphics[width=\linewidth]{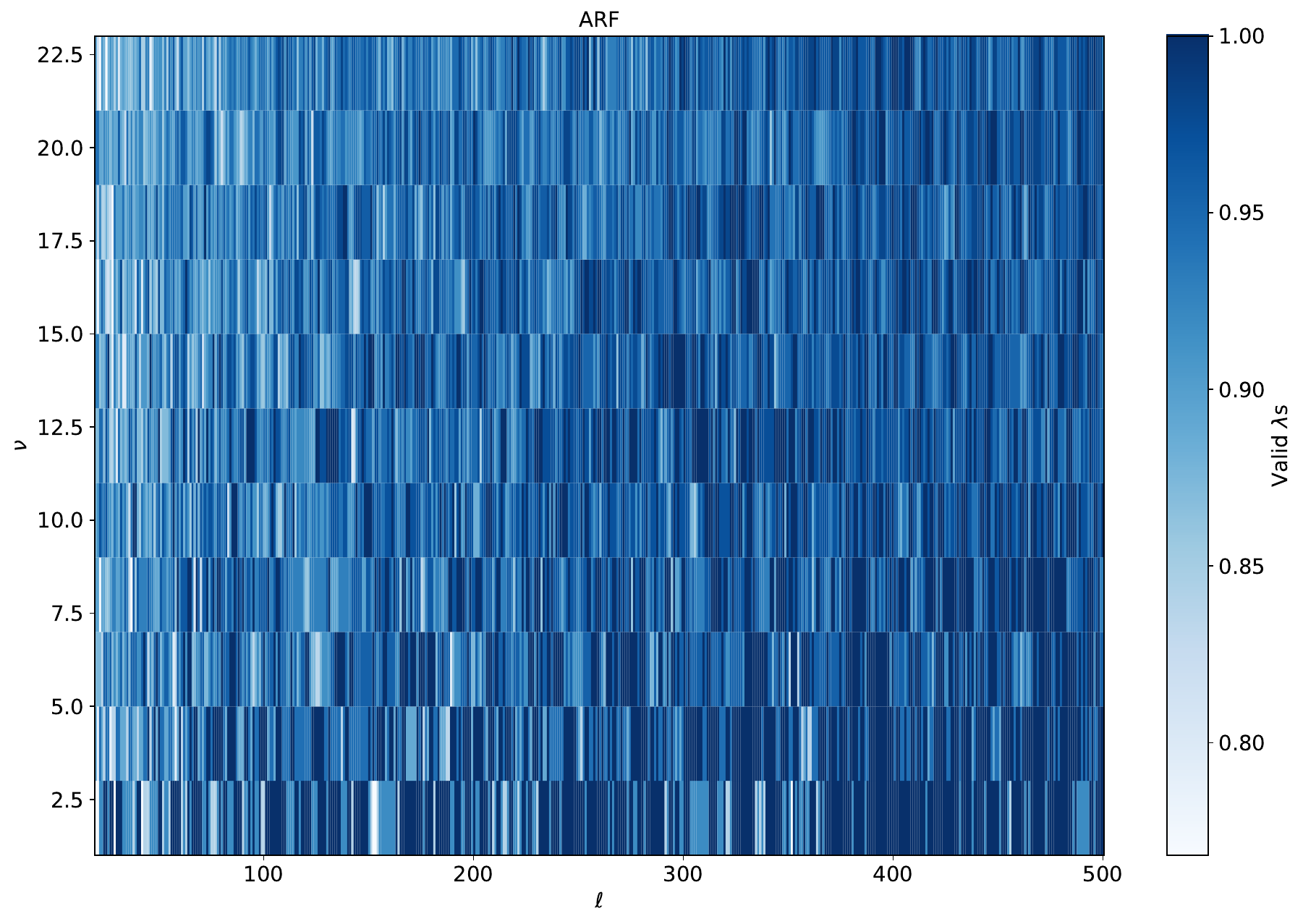}
        \caption{ARF valid eigenvector fraction versus $\nu$ and $\ell$.}
        \label{fig:ell2}
    \end{subfigure}
    
    \begin{subfigure}[b]{.55\linewidth}
        \includegraphics[width=\linewidth]{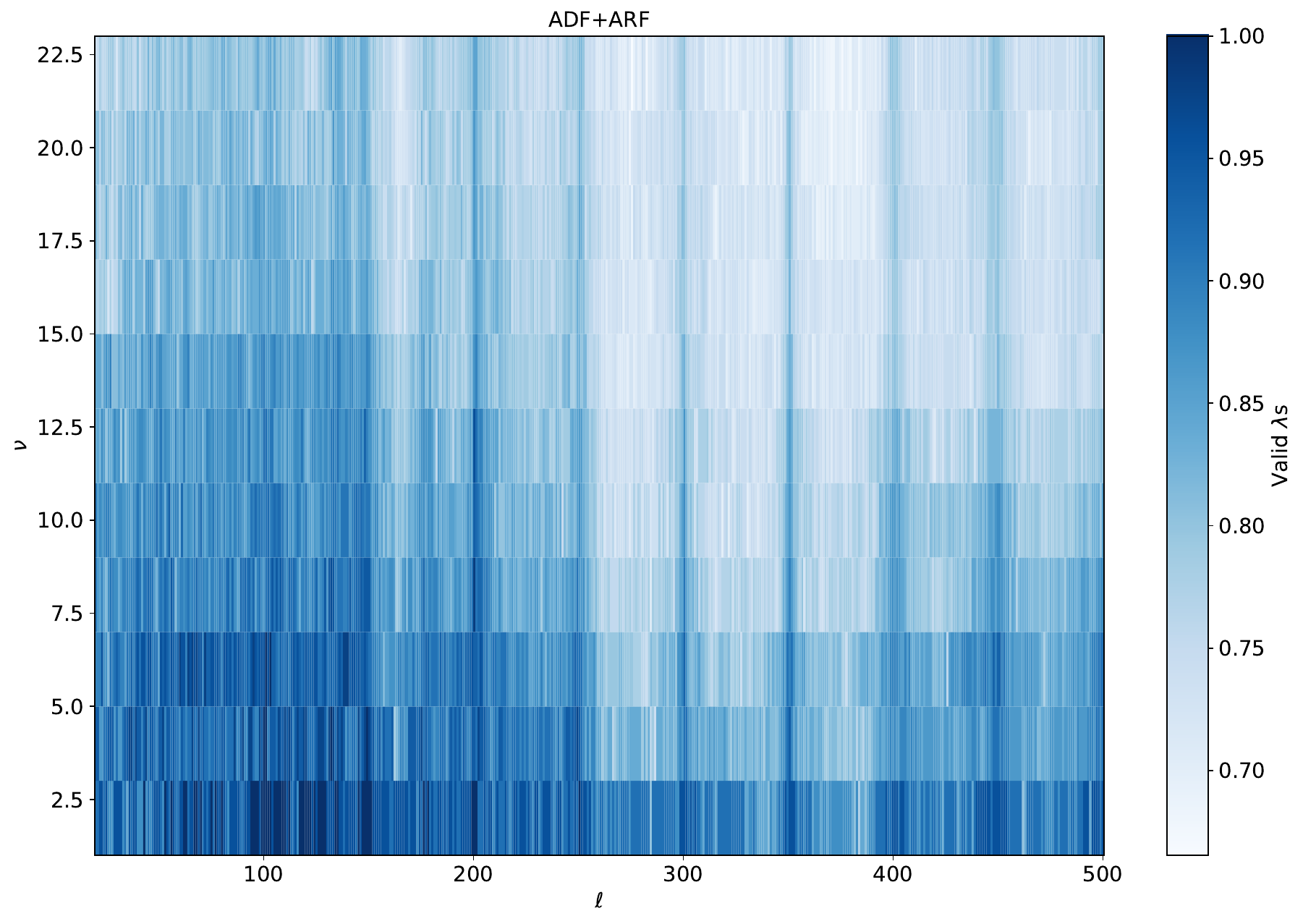}
        \caption{ADF+ARF valid eigenvector fraction versus $\nu$ and $\ell$.}
        \label{fig:ell3}
    \end{subfigure}
    
    \caption{Valid $\lambda$s fraction when varying $\nu$ and $\ell$. The darker blue represents higher fractions and lighter blue less valid $\lambda$s.}
    \label{fig:valid_ells}
\end{figure}

\begin{figure}
    \centering
    \includegraphics[width=0.5\linewidth]{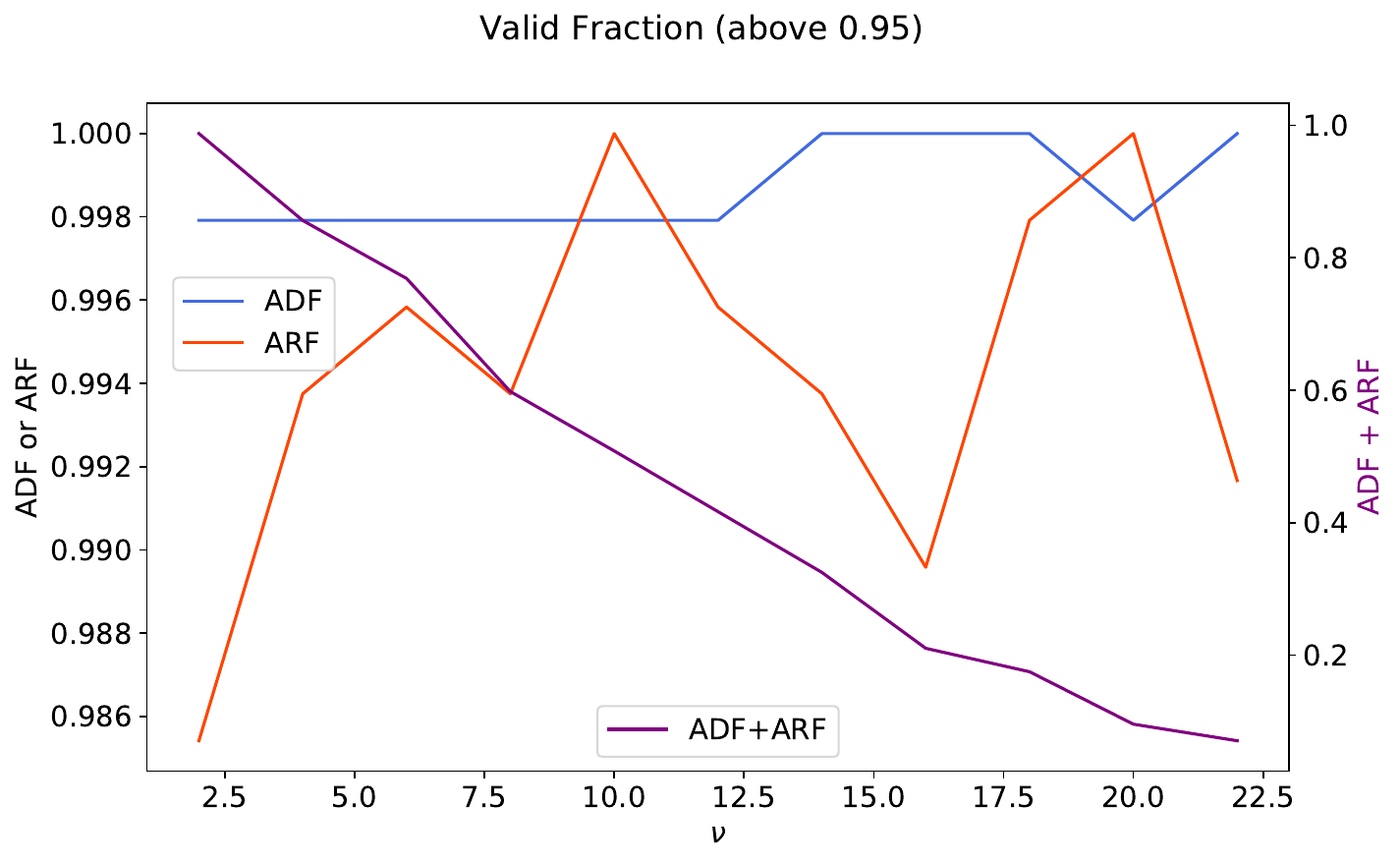}
    \caption{Fraction of Valid $\ell$s above 95\% of the total eigenvalues as a function of the number of neighbours.}
    \label{fig:valid_nu}
\end{figure}

To gain further insight on this, Fig.~\ref{fig:valid_ells} shows the fraction of surviving eigenvectors per $\ell$ for the ADF, ARF, and ADF+ARF probe setups in a {\it Euclid}-like configuration. Vertical axes represent neighbour shells included in the data vector, while horizontal axes refer to the $\ell$ multipole. The colour bar indicates the average fraction of surviving eigenmodes (according to the selection rules outlined above) relative to the initial number of eigenmodes corresponding to each cell in these 2D panels. These show that the fraction of surviving eigenvectors decreases for ARF, and yet further for ADF+ARF, suggesting that information is exhausted as more neighbours and probes are included in the tomographic analysis. In figure \ref{fig:valid_nu}, we show the percentage of valid eigenvalues that were accepted into the fisher matrix for different $\nu$, ADF+ARF is more unstable than the other cases as the number of neighbours increases in agreement to figure~\ref{fig:valid_ells}.

\acknowledgments

The authors would like to warmly acknowledge useful discussions with Jorge Martín-Camalich and José Ramón Bermejo-Climent. 
This work made use of the Milliways HPC computer located at the Instituto de Física in the Universidade Federal do Rio de Janeiro, managed and funded by \href{https://sites.google.com/view/arcos-ufrj/about?authuser=0}{ARCOS} (Astrophysics, Relativity and COSmology research group). PSF thanks Brazilian funding agency CNPq for PhD scholarship GD 140580/2021-2. PSF thanks Brazilian funding agency CAPES for the visiting scholarship at the IAC through CAPES-PRINT (PRINT: PROGRAMA INSTITUCIONAL DE INTERNACIONALIZAÇÃO) (grant number 88887.900701/2023-00). RRRR thanks CNPq for partial financial support (grant no. 309868/2021 - 1). C.H.-M. acknowledges the support of the Spanish Ministry of Science and Innovation via project PID2021-126616NB-I00, and the contribution from the IAC High-Performance Computing support team and hardware facilities.

\newpage
\bibliographystyle{JHEP}
\bibliography{biblio.bib}

\end{document}